\documentclass[12pt]{article}
\usepackage{graphicx}
\usepackage[utf8]{inputenc}
\usepackage{amsmath}
\usepackage{amsfonts}
\usepackage{amssymb}
\usepackage{graphicx}
\usepackage{color}

\usepackage{hyperref}
\hypersetup{
	hidelinks
}

\DeclareMathOperator{\sign}{sign}

\usepackage[sorting=none,citestyle=numeric-comp]{biblatex}
\renewbibmacro{in:}{}

\newcommand{\comm}[1]{} 

\begin{document}

\begin{center}
{	\Large \bf 
			Perturbations of classical fields by gravitational shockwaves }
\end{center}

\bigskip
\bigskip

\begin{center}
		D.V. Fursaev$^\dag$~,~E.A.Davydov$^{\dag\ddag}$~,~I.G. Pirozhenko$^{\dag\ddag}$~,~V.A.Tainov$^{\dag\ddag}$
	\end{center}
	
	\bigskip
	\bigskip
	\date{today}

	\begin{center}
		{\dag \it  Bogoliubov Laboratory of Theoretical Physics\\
			Joint Institute for Nuclear Research\\
			141 980, Dubna, Moscow Region, Russia}\\
			
			\bigskip
			
			and\\
			
			\bigskip
			
		\ddag	{\it Dubna State University, 
			Universitetskaya st. 19\\ }
		\medskip
	\end{center}
	\bigskip

\bigskip
\bigskip

\begin{abstract}
Gravitational shockwaves are geometries where components of the transverse curvature have abrupt behaviour across null hypersurfaces, which are fronts of the waves. We develop a general approach to describe classical field theories on such geometries in a linearized approximation, by using free scalar fields as a model. Perturbations caused by shockwaves exist above the wave front and are solutions to a characteristic Cauchy problem with initial data on the wave front determined by a supertranslation of ingoing fields. A special attention is paid to perturbations of fields of point-like sources
generated by plane-fronted gravitational shockwaves. One has three effects: conversion of non-stationary perturbations into an outgoing radiation, a spherical scalar shockwave which appears  when the gravitational wave hits the source, and 
a plane scalar shockwave accompanying the initial gravitational wave. Our analysis is applicable to gravitational shockwaves of a general class including geometries sourced by null particles and null branes. 
\end{abstract}

\newpage

\section{Introduction}\label{intr}

As has been demonstrated in a series of publications \cite{Fursaev:2023oep,Fursaev:2023lxq,Fursaev:2022ayo}, straight null cosmic strings create nonstationary perturbations of classical fields of point-like sources.
Late-time asymptotics of these perturbations behave as spherical waves outgoing from the sources.  The effect holds for fields of point charges and point masses, 
where perturbations are electromagnetic and gravitational waves, respectively. Since gravitational field of a straight null string belongs to a class of shockwave
geometries,
one may suggest that the effect described above is more general and it holds for gravitational shockwaves created by different ultrarelativistic sources. In this work we show that this is indeed the case, at least for plane-fronted shockwaves.

Gravitational shockwaves are solutions to the Einstein equations. They can be created by different sources which move with the speed 
of light, for example, by
massless particles \cite{Dray:1985173} or null branes \cite{Zheltukhin:1988rys}.  In the impulsive limit the metric components have delta-function-like singularities on the wave front. The shockwave geometries are characterized by abrupt changes of the transverse curvature tensor 
on the wave fronts (in general, but not in case of null cosmic strings \cite{Fursaev:2017aap}).

There are different reasons for a sustainable interest to gravitational shockwaves. One is in their peculiar mathematical structure.
There is a class of geometries, known as pp-waves, where the impulsive limit is well defined despite the fact
that the Einstein equations are essentially non-linear. Fronts of the shockwaves are null hypersurfaces which become singular hypersurfaces in the impulsive limit.  Construction of space-times with singular null hypersurfaces 
has been suggested long ago by R.Penrose \cite{Penrose:1972xrn}.   Wave front singularities are not an obstacle 
to determine ``surface physics'' on the fronts, as has been first shown  by Israel and Barrabes  \cite{Israel:1966rt,Barrabes:1991ng}. 

Another reason of a special attention to shockwaves is related to celebrated observation by 't~Hooft \cite{tHooft:1987vrq}
on the so called graviton dominance 
in ultra-high-energy scattering. Classical scattering
on gravitational shockwaves reproduces contribution of all ladder diagrams of graviton exchanges in quantum amplitudes \cite{Kabat:1992tb}, see for a review \cite{DiVecchia:2023frv}.  Colliding shockwaves can be used to model creation of microscopical black holes in collisions of particles at Planckian energies.

In the present paper we consider effects of plane-fronted shockwaves on classical fields.  The core of our analysis is a method 
developed on the base of results  of \cite{Fursaev:2023oep, Fursaev:2023lxq, Fursaev:2022ayo} for null strings. As has been shown there one can find perturbations of classical fields, obeyed by relativistic second order equations, by solving characteristic Cauchy problem.
The corresponding  initial data are set on the string event horizon, which is the shockwave front if one describes null string geometry in terms of shockwaves.

The results of this work are twofold. First, we show that perturbations of classical fields, caused by quite general type of 
gravitational plane-fronted 
shockwaves, can be found by solving characteristic Cauchy problem with initial data set on shockwave fronts. In a linear 
approximation these initial data are Lie derivatives of incoming field generated by supertranslations. Second, we apply our method to study
perturbations of fields of point-like sources crossing the shockwave fronts. We do all calculations by using 
a scalar field theory to model effects of the shockwaves on the Coulomb  field of a point charge or a Newton potential of a point mass.
The perturbations are shown to behave at late-time as outgoing spherical waves. For example,
the shockwave from a massless particle creates a perturbation of the field of a point scalar charge 
which looks at large times as
\begin{equation}\label{i.1}
	\phi_{sw}(r,U,\Omega) \simeq {\Phi(U,\Omega) \over r} + O(r^{-2})~~.
\end{equation}
Here $\Omega$ are spherical coordinates, $U$ is a retarded time, and (\ref{i.1}) implies that $r\gg U$. Since amplitude $\Phi$ depends on time there is an energy flux to the future null infinity.

In this regard one may expect that perturbations generated by gravitational shockwaves have precisely the same properties as in the case of null string geometries. This is not quite true: we demonstrate that gravitational shockwaves induce shockwaves in the field system itself. 
In the considered model there appears a plane-fronted scalar shockwave which propagates together with the initial gravitational wave. Also, when the gravitational wave hits the point-like source, it creates another 
outgoing scalar shock. The reason for scalar shockwaves is in jumps of normal derivatives of $\phi_{sw}$ on the corresponding wave fronts.

After \cite{tHooft:1987vrq} studying quantum scalar fields in shockwave backgrounds has been a subject of many publications, see e.g.  \cite{Klimcik:1988az,Battista:2014mka,Lodone:2009qe,Constantinou:2011ju,Gray:2021dfk,Cho:2023dnf}. For description 
of quantum fields one needs to know a modification of single-particle modes under the action of a shockwave. Technically this is reduced 
to solving a characteristic Cauchy problem with a particular type of initial data determined by monochromatic waves.
The approach to field theories on shockwave backgrounds, which is discussed in the present work, is more general. It allows one to 
study perturbations of fields with different spins and in the presence of external sources. Also our focus is on 
potentially observable macroscopic physical effects behind the wave fronts rather than on quantum effects at Planckian energies.

The paper is organized as follows. An introduction
for plane-fronted shockwaves travelling in Minkowski space-time is presented in Section \ref{S1}, along with
shockwaves with time-dependent profiles and definitions of surface physics on wave fronts.
In Section  \ref{S2} we describe properties of classical fields on shockwave backgrounds.
We start with engineering shockwave geometries based on the Penrose supertranslations in Section  \ref{S2.1}.
Soldering procedure of \cite{Blau:2015nee} and the Carroll nature of the supertranslations are discussed in Section  \ref{S2.2}.
This allows us to define soldering conditions for classical matter fields on wave fronts, see Section  \ref{S2.3},
where the key role is played by the Carroll variations of fields.  Soldering conditions for a point-like scalar source 
and the scalar field model we use are introduced in Sections \ref{CD} and \ref{SM}.
For chosen soldering conditions the field is continuous across the wave front but tangent components of the stress-energy tensor of 
field have abrupt behaviour, see Section \ref{ND}. Thus, there is a plane-fronted shockwave of the scalar field itself.
Perturbations of fields for point-like sources are calculated in Section \ref{S3}. As a technical trick, which is useful in a linearized 
approximation, we consider a shockwave of a general profile as a Fourier superposition of shockwaves with ``elementary'' profiles. 
Perturbations of ``elementary''  shockwaves are easier to study. A useful integral representation of the perturbation, which is a 
starting point of our analysis, is found in Section \ref{S3.1}.  It is shown in Section \ref{S3.2} that the gravitational shockwave, when it hits the source, produces the second shockwave in the field system, a spherical outgoing wave. In Section \ref{S3.3} we derive late-time asymptotic of the perturbation for 
shockwaves with ``elementary'' profiles, and check in Section \ref{S3.4} that with its help one reproduces results earlier known 
for null-string geometries. Section \ref{S4} considers perturbations of the field of the point-like source caused by shockwaves from null particles,  
Section \ref{S4.1}, and null branes, Section \ref{S4.2}. In both cases we find asymptotics of the perturbations in an analytic form. 
A peculiar property of null branes is that the perturbation vanishes inside the spherical shock produced when the brane hits the source.
We finish in Section \ref{S5} with final comments and a summary  of the results. Technical details related to the analysis
of Section \ref{S3.1},\ref{S3.2} 
are left for Appendix \ref{App1}.

\section{Plane-fronted shockwaves and their sources}\label{S1}
\setcounter{equation}0
\subsection{PP-waves}\label{S1.1}

We start with a very short introduction to description of plane-fronted shockwaves which are considered in the present work.
The important example are pp-wave geometries 
introduced by Ehlers and Kundt,  
\begin{equation}\label{1.1}
ds^2=-dv du -H(u,y)du^2+d y_{\perp}^2~~,
\end{equation}
where $v=t+x$,  $u=t-x$ are advanced and retarded null coordinates and $H(u,y)$ are some functions with a finite support along
$u$.  
Metrics
(\ref{1.1}) are plane-fronted waves with parallel propagation along the $x$ axis, from the left to the right. Outside
the plane wave, where $H(u,y)=0$,  the space-time is Minkowskian. 
In (\ref{1.1})  notation $d y_{\perp}^2\equiv\sum_i dy_i^2$ is used for the part of the metric which depends on coordinates $y_i$ orthogonal to the direction of motion of the wave.  We assume in what follows that $i=1,2$.

A peculiar property of (\ref{1.1})  is that the corresponding Ricci tensor has the only no-vanishing component 
\begin{equation}\label{1.2}
R_{uu}=\frac 12 \partial^2_{\perp} H~~.
\end{equation}
The right hand side of  (\ref{1.2}) is linear in $H$ despite that components of the curvature tensor are non-linear in metric.
This fact allows one to consider a particular case of pp-waves
\begin{equation}\label{1.3}
ds^2=-dv du -f(y)\chi(u)du^2+d y_{\perp}^2~~,
\end{equation}
where one can take the so called impulsive limit
\begin{equation}\label{1.4}
\chi(u)~\to~\delta(u)~~.
\end{equation}
According to (\ref{1.2})  such geometries are sourced by the stress-energy tensor with the only nontrivial 
component
\begin{equation}\label{1.5}
T_{uu}=\delta(u) \sigma(y)~~,~~ \sigma(y)=\frac{1}{16\pi G}~\partial^2_{\perp}f(y)~~,
\end{equation}
which corresponds to a massless source with the trajectory $u=0$. ($G$ is the Newton coupling). 
The source is distributed in a plane orthogonal to the direction of
motion with the surface energy density $\sigma(y)$. The front of the impulsive pp-wave is the null hypersurface $u=0$. Since
the source and the front move with the speed of light, the trajectory of the source lies on the front.

Some examples of sources which create impulsive pp-waves are listed in Table \ref{t1}, see \cite{Lousto:1990wn}.
They are obtained by the Aichelburg-Sexl boosts   of a point massive particle~\cite{Aichelburg:1970dh}  or  known topological defects such as cosmic string, domain wall~\cite{Vilenkin:1981zs}, and global monopole~\cite{Barriola:1989hx}. The parameter $E$ denotes the energy of a massless   particle $E_p$, the energy per unit length of the null string $E_s$, 
and the energy per unit area of the domain wall $E_{w}$. The string is supposed to be stretched along $y_2$ axis and 
$\rho\equiv (y_{\perp}^2)^{1/2}$.
\begin{table}	
	\begin{center}\label{t1}
		
		\begin{tabular}{lll}
			\hline
				Ultrarelativistic  source & $\sigma(y)$  & $f(y)$ \\
			\hline
			Massless point particle   & $E_p\delta(\rho)$  & $8 G E_{p} \ln \rho $ \\
			Global monopole  & $E_{m}/\rho$ & $16\pi G E_{m}\rho $ \\
			Null straight cosmic string  & $E_{s}\delta(y_1)$ & $ 8\pi G E_s |y_1| $ \\
			Domain wall & $E_{w}$ & $4\pi G E_{w} \rho^2 $  \\
			\hline
		\end{tabular}
	\end{center}
	\caption{Examples of ultrarelativistic  sources  with two-dimensional energy distribution $\sigma(y_{\perp})$  considered in \cite{Lousto:1990wn}. The parameters $E_p$, $E_m$, $E_s$, $E_w$ are energies of the sources. }
\end{table}

\subsection{Plane-fronted waves with time-dependent profiles}\label{S1.2}

In our work we also consider a more general class of shockwave geometries, which are not pp-waves, by allowing  
time dependence of the profile functions. The corresponding metrics and nonvanishing 
components of the Riemann and Ricci tensors are
\begin{equation}\label{1.6}
ds^2=-dv du -H(v,u,y)du^2+d y_{\perp}^2~~,
\end{equation}
\begin{equation}\label{1.7}
R_{uaub}=\frac 12 \partial_a\partial_b H~~,
\end{equation}
\begin{equation}\label{1.8}
R_{uu}=\frac 12 \partial^2_{\perp} H+2H\partial_v^2H~~,~~R_{ua}=\partial_v\partial_a H~~.
\end{equation}
Here and in what follows we use the convention that indices $a,b$ numerate coordinates $v, y_{\perp}$, while $i,j$
correspond only to $y_{\perp}$.

Theory is linear, if $\partial_v^2H=0$, that is when the profile functions are linear functions of time.
The impulsive limit, $\chi(u)\to \delta(u)$, is well-defined if
\begin{equation}\label{1.9}
H(u,y,v)=\chi(u)f(y,v)~~,~~\partial_v^2f=0~~.
\end{equation}
A discussion of this condition for shockwaves in arbitrary space-time can be found in \cite{Sfetsos:1994xa}.

We admit arbitrary profile functions $f(y,v)$ by restricting computations to effects which are linear in $f$.
This allows us to consider plane shockwaves with most general surface properties
of the wave fronts.  In the impulsive limit and in the linear approximation with respect to $f$
components of the Ricci tensor (\ref{1.8}) become
\begin{equation}\label{1.11}
R_{uu}\simeq \frac 12 \delta(u)\partial^2_{\perp}f~~,~~R_{ua}=\delta(u)\partial_v\partial_a f~~,
\end{equation}
which correspond to the stress-energy tensor of the source
\begin{equation}\label{1.19a}	
T_{\mu\nu}=\delta(u)\left (\sigma l_\mu l_\nu+j_i ( l_\mu e_\nu^i+ l_\nu e_\mu^i)+p\delta_{ij}e_\mu^ie_\nu^j\right)~~,
\end{equation}
\begin{equation}\label{1.19b}	
\sigma={ f_{,ii}\over 16\pi G}~~,~~j_i=-{f_{,vi} \over 8\pi G}~~,~~p={f_{,vv} \over 4\pi G}~~,
\end{equation}
where $l=2\partial_v$, $e^i=\partial_i$ are tangent vectors to $\cal N$.  Surface energy in (\ref{1.19b}) reproduces definitions from the previous Section. The appearance of nontrivial surface  current $j_i$ and surface pressure $p$ is the consequence of time dependence of the 
profile function. The fact that energy depends on advanced time implies the surface continuity relations
\begin{equation}\label{1.20}	
2\partial_v\sigma=-\partial^i j_i~~,~~2\partial_vj_i=-\partial_i p~~.
\end{equation}

We complete this Section with requirements for the surface
stress-energy tensor (\ref{1.19a}).  One needs (\ref{1.19a}) to satisfy
at least the weak energy condition  
\begin{equation}\label{1.23}	
T_{\mu\nu} u^\mu u^\nu \geq 0~~
\end{equation}
for any time-like vector $u$. The necessary condition for (\ref{1.23}) is that
the following  quadratic form is  positive-definite:
\begin{equation}\label{1.24}	
A_{ab}x^a x^b \geq 0~~,
\end{equation}
where $A_{vv}=\sigma$,  $A_{vi}=j_i$,  $A_{ij}=p\delta_{ij}$. 
Eigenvalues of matrix $A_{ab}$ are 
\begin{equation}\label{1.25}	
\lambda=p~~,~~\lambda_{\pm}=\frac 12
\left(p+\sigma\pm \sqrt{(p-\sigma)^2+4j^2} \right)~~,
\end{equation}
where $j^2=\delta_{ij}j^i j^j$. Condition
(\ref{1.24}) is guaranteed if the eigenvalues are non-negative.

If $p>0$ the fact that $ \lambda_->0$ implies
\begin{equation}\label{1.26}	
p\sigma -j^2 \geq 0~~.
\end{equation}
Two restrictions, (\ref{1.26}) and $p>0$, are necessary conditions for the weak energy condition.
Their consequence is that $\sigma>0$. Also  $j_i=0$, if $\sigma=0$ or $p=0$.  It means, for example, that one cannot make time-dependent the energy parameter
$E$ in case of null strings and null particles, see Table \ref{t1}.

The impulsive limit is well-defined if $f$ is linear in $v$, see  (\ref{1.9}). According to (\ref{1.19b}), in this case, $p=0$.
Then (\ref{1.26}) requires that $j_i=0$, which is possible only for spatially constant profiles $f$ for which $\sigma=0$.

\section{Classical fields and shockwaves }\label{S2}
\setcounter{equation}0

\subsection{Engineering shockwave geometries}\label{S2.1}

\begin{figure}
	\begin{center}
		\includegraphics[height=5cm]{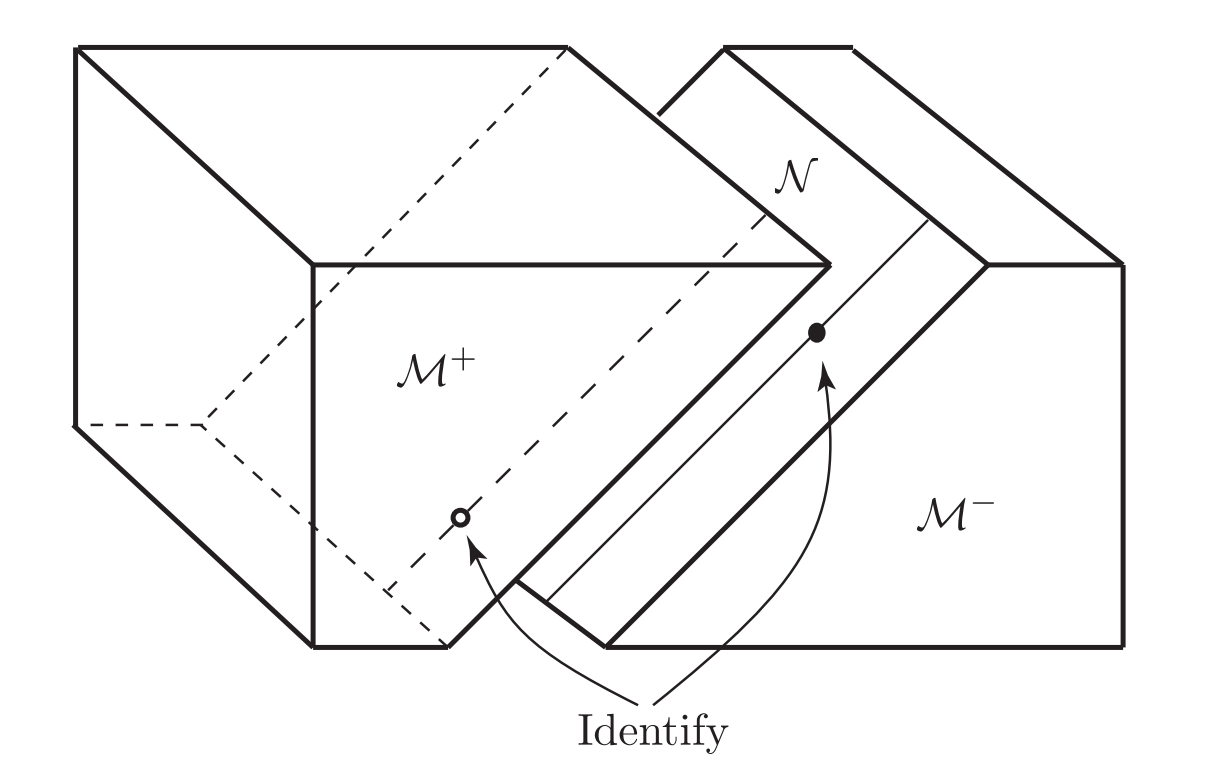}
		\end{center}
	\caption{Constructing shockwaves by soldering manifolds ${\cal M}^+$ and ${\cal M}^-$ along a null hypersurface
$\cal N$. This picture is taken from \cite{Podolsky:2017pth}.}
	\label{fig1}	
\end{figure}

In what follows we do not use (\ref{1.3}) to calculate the impulsive limit of shockwaves with arbitrary profiles. We employ an alternative method described below.
The procedure to construct a space-time which is locally Minkowski $R^{1,3}$ except a null hypersurface where the Riemann tensor
has singular structure (\ref{1.7}) is based on the Penrose approach \cite{Penrose:1972xrn}. 
One cuts $R^{1,3}$ by a null hypersurface $u=0$, which we also denote by $\cal N$, into two 'halfs', ${\cal M}^+$ ($u>0$) and 
${\cal M}^-$ ($u<0$). Then ${\cal M}^\pm$ are ``glued'' again along $\cal N$ with the following identification of points:
\begin{equation}\label{1.10}
v_+= v-f(v,y)~~,~~y^i_+=y^i~~,
\end{equation}
see Figure \ref{fig1}.
Transformations (\ref{1.10}) are called supertranslations. They leave invariant metric on $\cal N$ and, thus, make an infinite 
dimensional group of isometries of $\cal N$, also known as Carroll transformations 
\cite{Duval:2014lpa,Duval:2014uoa,Ciambelli:2023tzb,Ciambelli:2023xqk}.
 
Such a behaviour of the Ricci tensor is related to jumps of first derivatives of the metric on the fronts of shockwaves. 
With the help of  a null vector $n$ at $\cal N$, such as $(n\cdot l)=-1$, 
$(n\cdot e_i)=0$ one can define the transverse curvature 
\begin{equation}\label{1.11b}
{\cal C}_{ab}=n_{\mu;\nu} e_a^\mu  e_b^\nu~~,
\end{equation}
where $e_v=l, e_i=e^i$.  Then 
\begin{equation}\label{1.21}
\sigma={1 \over 8\pi G}[{\cal C}_{ii}]~~,~~j_i=-{1 \over 8\pi G}[{\cal C}_{vi}]~~,~~p={1 \over 8\pi G}[{\cal C}_{vv}]~~,
\end{equation}
\begin{equation}\label{1.22}
[{\cal C}_{ab}]\equiv {\cal C}_{ab}^+-{\cal C}_{ab}^-~~,
\end{equation}
where ${\cal C}_{ab}^\pm$ correspond to ${\cal M}^\pm$.  The tensor structure ${\cal C}_{ab}$
has been introduced in \cite{Poisson:2002nv}  following \cite{Israel:1966rt,Barrabes:1991ng}. For other equivalent definitions
discussed latter see, e.g., \cite{Aghapour:2018icu, Jafari:2019bpw}. 
In the considered case 
\begin{equation}
	[{\cal C}_{vv}]=2[\partial_u g_{vv}]~~,~~ 
	[{\cal C}_{vi}]=[\partial_u g_{vi}]~~,~~
	[{\cal C}_{ii}]=\frac{1}{2}[\partial_u g_{ii}]~~.
\end{equation}
To see how supertranslations (\ref{1.10}) lead to (\ref{1.11}) one needs to know $\partial_u g_{\mu\nu}^{~+}$
on $\cal N$.  In the linear approximation this can be easily done by using a soldering 
method of \cite{Blau:2015nee}.
Coordinate transformations of ${\cal M}^+$ in a neighborhood of $\cal N$ are
\begin{equation}\label{1.12}
x^\mu_+=x^\mu-\zeta^\mu(x)~~.
\end{equation}
Since (\ref{1.12}) should not change the metric on $\cal N$ one requires that
\begin{equation}\label{1.15}
{\cal L}_\zeta g_{\mu\nu}^{~-}|_{u=0}=(\zeta_{\mu,\nu}+\zeta_{\nu,\mu})|_{u=0}=0~~,
\end{equation}
which immediately yields:
\begin{equation}\label{1.13}
\zeta^\mu(x)\simeq f\delta^\mu_v+\frac u2 \eta^{\mu a}f_{,a}~~,~~
\end{equation}
\begin{equation}\label{1.16}	
u^+=u(1+f_{,v})~~,~~v^+=v-f~~,~~y_i^+=y_i-u f_{,i}/2~~.
\end{equation}
In vicinity of $\cal N$ deviation of the metric from the flat one is nontrivial
\begin{equation}\label{1.17a}	
g_{\mu\nu}^+ \simeq g_{\mu\nu}^-+{\cal L}_\zeta g_{\mu\nu}^{~-}=
\eta_{\mu\nu}+\zeta_{\mu,\nu}+\zeta_{\nu,\mu}~~,
\end{equation}
which yields
\begin{equation}\label{1.17}	
\partial_u g_{v v}^{~+}(u,{\bf x})|_{u=0}=f_{,vv}~~,~~ 	
\partial_u g_{v i}^{~+}(u,{\bf x})|_{u=0}=f_{,vi}~~,~~ 	
\partial_u g_{ij}^{~+}(u,{\bf x})|_{u=0}=f_{,ij}~~, ~~{\bf x}={v,y^i}~~.
\end{equation}
Substitution of (\ref{1.17}) in  (\ref{1.21})  correctly reproduces (\ref{1.19b}). We call (\ref{1.16}) soldering transformations.

There is an ambiguity in definition of the profile function $f$: one can add to the r.h.s. of \eqref{1.12} any Killing vector field of Minkowski space-time. That is one can apply Poincar\'{e} transformations which do not change physical characteristics of the shockwave. We will use this fact in what follows. 

To summarize, the approach by Penrose, accompanied by the soldering prescription (\ref{1.15}), allows one to reproduce
shockwave geometries in the impulsive limit without appealing to the shockwave metric. Supertranslations generate shockwaves with different profiles.

\subsection{Soldering transformations and Carroll transformations}\label{S2.2}

Before we proceed some comments are in order. One can introduce on $\cal N$ vector fields $V=V^a \partial_a$
and one-forms  $\theta=\theta_a d x^a$, where $x^a=(v,y^i)$.  Note that index $a$ cannot be risen or 
lowered since the metric of  $\cal N$ is degenerate.

The Carroll symmetries (supertranslations)
are generated by the vector field on $\cal N$
\begin{equation}\label{2a.3}
\bar{\zeta}^a({\bf x})=f({\bf x})\delta^a_v~~,~~{\bf x}={v,y^i}~~.
\end{equation}
The supertranslations induce 
transformations of components of $\theta_a$, $V^a$ and make different representations of the Carroll group.
They leave invariant forms like $\theta_a V^a$.
One can also define direct products of the representations by considering tensor structures on $\cal N$
with arbitrary numbers of upper and lower indices $a,b,...c$. 

Soldering transformations near (\ref{1.13}) have the important property: they map tangent (cotangent) spaces
of $\cal N$ to themselves. Indeed, transformations of 4-dimensional one-forms $\theta_\mu$ and vectors $V^\mu$ on ${\cal M}^+$ 
 are determined by the Lie derivatives,
\begin{equation}\label{2a.6}
{\cal L}_\zeta V^\mu=\zeta^\nu \partial_\nu V^\mu-V^\nu \partial_\nu \zeta^\mu~~,
\end{equation}
\begin{equation}\label{2a.7}
{\cal L}_\zeta \theta_\mu=\zeta^\nu \partial_\nu \theta_\mu+\theta_\nu \partial_\mu \zeta^\nu~~.
\end{equation}
At $\cal N$ these quantities have the following components:
\begin{equation}\label{2a.4}
{\cal L}_\zeta V^a=\hat{\cal L}_{\bar{\zeta}} V^a-\frac 12 \eta^{ab}f_{,b} V^u~~,~~{\cal L}_\zeta V^u=\partial_v(f V^u)~~,
\end{equation}
\begin{equation}\label{2a.5}
{\cal L}_\zeta \theta_a=\hat{\cal L}_{\bar{\zeta}} \theta_a~~,~~{\cal L}_\zeta \theta_u=f \theta_{u,v}-f_{,v}\theta_u+\frac 12 f_{,i}\theta_i~~,
\end{equation}
where $\hat{\cal L}_{\bar{\zeta}}$ is the Lie derivatives on $\cal N$ generated by $\bar{\zeta}$.  Thus,
\begin{equation}\label{2a.8}
{\cal L}_\zeta V|_{\cal N}=\hat{\cal L}_{\bar{\zeta}}V~~,~~\mbox{if}~~(n\cdot V)|_{\cal N}=0~~,
\end{equation}
\begin{equation}\label{2a.9}
{\cal L}_\zeta \theta|_{\cal N}=\hat{\cal L}_{\bar{\zeta}}\theta~~,~~\mbox{if}~~(l\cdot \theta)|_{\cal N}=0~~,
\end{equation}
where null vectors $l$ and $n$ at $\cal N$ have been introduced in the previous Section. 

There is an exceptional case when soldering condition (\ref{1.15}) holds globally, that is, soldering transformations are
isometries of ${\cal M}^+$. This is the case of so called null rotations. The corresponding shockwave geometries are 
created by null cosmic strings \cite{Fursaev:2017aap}.

\subsection{Soldering of fields on wave fronts}\label{S2.3}

Since ${\cal M}^+$ near $\cal N$ is considered as coordinate transformation of ${\cal M}^-$ generated by the vector field 
$\zeta^\mu$, defined by (\ref{1.13}),
we are forced to require that tangent and fibre bundles on ${\cal M}^+$ near $\cal N$ are  corresponding Lie-transformed structures 
on  ${\cal M}^-$. This guarantees a continuous (but not smooth) soldering of these structures across $\cal N$.
The soldering conditions for the metric and fields $\phi$, thus, look as 
\begin{equation}\label{1.12a}
x^+=x^--\zeta~~,~~u=0~~,
\end{equation}
\begin{equation}\label{2a.10}
g^+_{\mu\nu}=g^-_{\mu\nu}+{\cal L}_\zeta g^-_{\mu\nu}=g^-_{\mu\nu}~~,~~
u=0~~,
\end{equation}
\begin{equation}\label{2a.11}
\phi^+=\phi^- +{\cal L}_\zeta \phi^-~~,~~u=0~~.
\end{equation}
Here notation $\phi$ in (\ref{2a.11}) stands for fields of different spins. The indexes are implied but not shown  explicitly.

Keeping in mind the discontinuity of the transverse curvature (\ref{1.22}) we require that (\ref{2a.11}) holds for fields but may be violated for normal derivatives of fields. In Section \ref{ND} we illustrate this discontinuity for scalar models.

Equation (\ref{2a.11}) is a starting point to find perturbations of fields caused by shockwaves.  First of all one is interested in free fields which obey in $R^{1,3}$ the equations:
\begin{equation}\label{2.1}
P \phi(x)=j(x)~~.
\end{equation}
For integer spin fields $P$ is some relativistic second order hyperbolic differential operator, and $j(x)$ is a source. 
The operator $P$ may be degenerate, as for gauge fields and gravitons. We describe a general approach how to find perturbations to  (\ref{2.1}) and demonstrate its explicit realization in case of scalar fields.

In case of the gravitational shockwave one has two problems like (\ref{2.1}), in  ${\cal M}^-$ and  in ${\cal M}^+$,
\begin{equation}\label{2.2}
P\phi^\pm =j^\pm~~,
\end{equation}
\begin{equation}\label{2.3}
\phi^+=\phi^- +{\cal L}_\zeta \phi^-~~,~~u=0~~.
\end{equation}
\begin{equation}\label{2.4}
j^+=j^- +{\cal L}_\zeta j^-~~,~~u=0~~.
\end{equation}
Note that the source is also affected by the shockwave. Usually the source
is related to a flow of freely moving matter which is subject to the gravitational memory effect after crossing the shockwave front~\cite{Shore:2018kmt}.  It is this type of the sources we consider in this paper.

If $P$ is degenerate one has to impose gauge-fixing conditions. As a result, 
soldering condition (\ref{2.3}) should be considered only for some components of the field. In the case of null string space-times soldering  in the Maxwell theory and in the linearized  gravity has been
carried out  along these lines in \cite{Fursaev:2022ayo,Fursaev:2023lxq,Fursaev:2023oep}.
It has been shown there that for some Lorentz invariant gauge-fixing conditions one can fix only those parts of fields which are in the tangent space to $\cal N$.

Since $\cal N$ is also an event horizon of the wave, the fields in ${\cal M}^-$ are not affected by the wave.
A unique solution in ${\cal M}^-$ can be found by requiring  asymptotic conditions at the past infinities. 

The problem   in ${\cal M}^+$  is different. This region is causally connected to the shockwave,
and solution in ${\cal M}^+$ should obey given initial data on  $\cal N$.  One has here a characteristic Cauchy problem. Characteristic Cauchy problems are initial value problems with initial data on null hypersurfaces.
Perturbations $\phi_{sw}$, which are present in ${\cal M}^+$,  can be defined by comparing 
unperturbed solution $\phi$
with $\phi^+$ for the same source. That is,
\begin{equation}\label{2.5}
\phi_{sw}\equiv \phi^+ -\phi~~,~~j^+(x)\equiv j(x)~~,~~u>0~~.
\end{equation}
Hence, the two solutions, $\phi$
with $\phi^+$, obey the same 
inhomogeneous equations at $u>0$ but have different conditions at $\cal N$. This 
results in the following homogeneous characteristic problem for the perturbations:
\begin{equation}\label{2.6}
P\phi_{sw}=0~~,~~u>0~~,
\end{equation}
\begin{equation}\label{2.8}
\phi_{sw}=\phi^+-\phi= {\cal L}_\zeta \phi^--(\phi-\phi^-)~~,~~u=0~~.
\end{equation}
Equation (\ref{2.8})  is the consequence of (\ref{2.3}). In  ${\cal M}^-$ fields
$\phi$ and $\phi^-$ are solutions to problems with sources $j$ and $j^-$, respectively.
We are interested in the linear in $f$ approximation.  In this approximation, according to (\ref{2.4})
\begin{equation}\label{2.9}
j^-=j - {\cal L}_\zeta j~~,~~u=0~~,
\end{equation}
and initial conditions (\ref{2.8}) can be written as
\begin{equation}\label{2.10}
	\phi_{sw}=\phi^+-\phi= {\cal L}_\zeta \phi-\delta_j\phi~~,~~u=0~~.
\end{equation}
Here $\delta_j\phi$ is a variation of the solution under change $j$ to $j^-$. To determine 
$\delta_j\phi$ one needs information about the source.

\subsection{Initial data for the case of a  point source}\label{CD}

We consider now the case when $j$ corresponds to a point source which moves freely.
If $V$ is a four-velocity of the source in ${\cal M}^+$ and  $x_o$ are its coordinates on 
$\cal N$, then $j^-$ corresponds to a source with four-velocity and coordinates
on the wave front
\begin{equation}\label{2.11}
x_o^-=x_o+\zeta (x_o)~~,~~u=0~~,
\end{equation}
\begin{equation}\label{2.12}
V^-=V -{\cal L}_{\zeta(x_o)}V~~,~~u=0~~.
\end{equation}
Here we used (\ref{1.12a}). Without loss of the generality we put $x_o=0$ at $u=0$.

Since the source moves freely, its trajectory in ${\cal M}^-$ 
is obtained  under an action of the Poincar\'{e} group which shifts and boosts 
the initial trajectory. Let the corresponding generators be
\begin{equation}\label{2.13}
\zeta_o^\mu(x)=\zeta^\mu (x_o)+\omega^\mu_{~~\nu}(x^\nu-x_o^\nu)~~,
\end{equation}
where $\omega_{\mu\nu}=-\omega_{\nu\mu}$. Lorentz matrices $\omega^\mu_{~~\nu}$ can be found
by requiring that Carroll and Poincar\'{e} transformations coincide on $\cal N$ 
\begin{equation}\label{2.14}
{\cal L}_{\zeta_o}V={\cal L}_{\zeta(x_o)}V~~,~~u=0~~.
\end{equation}
If (\ref{2.14})  holds, $\phi^-$ is just the Poincar\'{e} transformation of the initial
field, $\phi^-=\phi-{\cal L}_{\zeta_o}\phi$,  and  initial conditions (\ref{2.10}) 
take the simple form:
\begin{equation}\label{2.15}
\phi_{sw}=({\cal L}_\zeta-{\cal L}_{\zeta_o})\phi={\cal L}_{\zeta'}\phi~~,~~u=0~~, 
\end{equation}
where $\zeta'=\zeta-\zeta_o$.
It implies, in particular, that $\phi_{sw}(x_o)=0$.  From (\ref{1.13}), (\ref{2a.6}),
(\ref{2.14}), after some algebra,  one finds the Lorentz part of
(\ref{2.13})
\begin{equation}\label{2.16}
\omega^\mu_{~~\nu}=	(\zeta_o)^\mu_{~~,\nu}=\zeta^\mu_{~~,\nu}(x_o)=
\delta^\mu_v f_{,\nu}(0)+\frac 12 \delta_\nu^u \eta^{\mu a} f_{,a}(0)~~.
\end{equation}
The important property of (\ref{2.15})
is that it allows one to determine the initial conditions for the perturbations solely
in terms of the unperturbed field $\phi$ and characteristics of the shockwave 
encoded in soldering transformation (\ref{1.13}). 

\subsection{Scalar model} \label{SM}

In the rest of the paper we study the model where $\phi$ is
a real massless scalar field and $P=\Box=-4\partial_u \partial_v+\partial_i^2$. We suppose that the source is located at the center 
of coordinates,
$j(x)=Q ~\delta^{(3)}(x,y^i)$, with $x=(v-u)/2$ being a spatial coordinate.
The solution to  (\ref{2.1})
is the potential of a point source with a scalar charge $Q$,
\begin{equation}\label{2.17}
\phi(x) = -\frac{Q}{4\pi\sqrt{x^2+y_i^2}}~~.
\end{equation}
We are interested in perturbations $\phi_{sw}$ of (\ref{2.17}) by a shockwave  with a profile function $f$.  According with the above analysis 
the characteristic Cauchy problem for  $\phi_{sw}$  in ${\cal M}^+$
is 
\begin{equation}\label{2.18}
\Box	\phi_{sw}(x)=0~~,~~u>0~~,
\end{equation}
\begin{equation}\label{2.19}
\phi_{sw}(x)=(\zeta^\mu(x)-\zeta_o^\mu(x))\partial_\mu\phi (x)~~,~~u=0~~.
\end{equation}
For our choice, $x_o=0$,
\begin{equation}\label{2.20}
\zeta_o^\mu(x)=\zeta^\mu (0)+\delta^\mu_v f_{,a}(0)x^a+\frac 12  \eta^{\mu a} f_{,a}(0)u~~,
\end{equation}
and condition (\ref{2.19}) takes on the final form:
\begin{equation}\label{2.21}
\phi_{sw}(x)=(f({\bf x}) -f(0)-f_{,a}(0)x^a)\partial_v\phi(x)~~,~~u=0~~.
\end{equation}
In next sections we study solutions to the problem (\ref{2.18}), (\ref{2.21}).

\subsection{Generated shockwaves of matter fields}\label{ND}

An immediate consequence of the chosen soldering conditions is that the gravitational shockwave generates a shockwave 
in the field system itself, a scalar shockwave which accompanies the initial gravitational wave. This effect 
is a result of discontinuities of the stress-energy tensor on $\mathcal{N}$ measured by an observer crossing the wave front.

An observer with four-velocity $V$ does not register discontinuities in field $\phi$ since observer's four-velocity transforms on $\mathcal{N}$ in accord with transformation of the field, see \eqref{2.3},
\begin{equation} \label{2.22}
	V_- \to V_+ = V_-+ \mathcal{L}_\zeta V_- |_\mathcal{N}~~.
\end{equation}
However other quantities, constructed from derivatives of $\phi$ may not be smooth on $\mathcal{N}$. We illustrate it for the quantity $\chi \equiv \partial_V \phi$ and show that $\chi$ does not transform similar to \eqref{2.3}. Namely, in general,
\begin{equation} \label{2.23}
	\Delta \chi = \Bigl[
					\chi_+ - \bigl(	\chi_- + \mathcal{L}_\zeta	\chi_- \bigr)
					\Bigr] 
					_\mathcal{N} \neq 0~~,
\end{equation}
where $\chi_\pm=\partial_{V_\pm} \phi_\pm$.
To proceed we represent $\Delta \chi $ as follows:
\begin{equation} \label{2.24}
	\Delta \chi  \simeq \Bigl[
						(V_+ \cdot \partial\phi_+) 
						- \bigl( 
						  (V_+ \cdot \partial\phi_-) 
						   + (V_+ \cdot \mathcal{L}_\zeta(\partial \phi_-)  )
						  \bigr)
						\Bigr] 
						_\mathcal{N}~~,
\end{equation}
where we used \eqref{2a.10}, the identity $\mathcal{L}_\zeta(\partial \phi) = \partial (\mathcal{L}_\zeta \phi)$, and the fact we are interested in the linear in $f$ approximation.

It is clear that $\Delta \chi = 0$, if $V_+$ is tangent to $\mathcal{N}$, since in this case
\begin{equation} \label{2.25}
	\bigl( V_+ \cdot \mathcal{L}_\zeta(\partial \phi_-) \bigr) |_\mathcal{N}
	=
	\bigl( V_+ \cdot \partial ( \mathcal{L}_\zeta \phi_- )|_\mathcal{N} \bigr) ~~.
\end{equation} 
Therefore, we need to investigate the case when $V$ is not tangent to $\mathcal{N}$.  If $V$ is replaced with vector $n$, where $n^2 =0, (n \cdot l)=-1, n \partial = \partial_u$ one has (see \eqref{2.24})
\begin{equation} \label{2.26}
	\Delta \chi  \simeq \Bigl[
						(n_+ \cdot \partial\phi_+) 
						-  (n_+ \cdot \partial)  (\phi_- +  \mathcal{L}_\zeta \phi_-) 
						\Bigr] 
						_\mathcal{N}~~.
\end{equation}
From \eqref{1.13} one finds:
\begin{equation} \label{2.27}
	(n_+ \cdot \partial)  (\phi_- +  \mathcal{L}_\zeta \phi_-)
	=
	\Bigl[
	(1-f_{,v})	\partial_u \phi_-
		+ f \partial_u \partial_v \phi_- + \frac{1}{2} f_{,i} \partial_i \phi_-
	\Bigr]_{u=0}~~.
\end{equation}
From now on we should specify the equations. As in Section \ref{SM}, we take $P = \Box = -4 \partial_u \partial_v + \partial_i^2$. In characteristic Cauchy problems normal derivatives $\partial_u \phi$ are not independent on fields $\phi$. Indeed,
\begin{equation} \label{2.28}
	\partial_u \phi_\pm (v,y)= \frac{1}{4} \int_{-\infty}^{v} dv' \Bigl[ \partial_i^2 \phi_\pm - j_\pm \Bigr]_{u=0}+F(y)~~.
\end{equation}
Here $F(y)$ is an arbitrary function, in general.  In our case $F=0$ since we require $\phi_\pm$ to decay at $v\to -\infty$.
By using \eqref{2.27}, \eqref{2.28} in \eqref{2.26}  we get the following result for the jump of normal derivatives on $\mathcal{N}$ :
\begin{equation} \label{2.29}
	\Delta \chi \simeq 
		\frac{1}{4} ( f_{,ii})\phi_+ 
		+ \frac{1}{4} f_{,v} \int_{-\infty}^{v} dv' \Bigl[ \partial_i^2 \phi_+ - j_+ \Bigr] 
		-\frac{1}{4} \int_{-\infty}^{v} dv' \Bigl[ \partial_i^2 \bigl( f_{,v'}\phi_+ \bigr) - f_{,v'} j_+ \Bigr]_{u=0}~~,
\end{equation}
for $\chi=\partial_u \phi$, which is expressed solely in terms of Cauchy data on $\mathcal{N}$ and the profile function $f$.
Equation \eqref{2.29} has a simple form in case when $f$ does not depend on time,
\begin{equation} \label{2.30}
	\Delta \chi \simeq 4\pi G \sigma \phi_+ ~~,
\end{equation}
where $\sigma$ is defined in \eqref{1.19b}. 

Discontinuities \eqref{2.29} and \eqref{2.30} are analogous to the transverse gravitational curvature (\ref{1.22}).
Let us define a tensor structure on $\cal N$
\begin{equation}\label{2.31}
\tau _{\mu\nu} \equiv \Bigl[T_{\mu\nu} [\phi_+]- (T_{\mu\nu} [\phi_-]+\mathcal{L}_\zeta T_{\mu\nu} [\phi_-])\Bigr] _\mathcal{N} ~~,
\end{equation}
by using the values of the stress-energy tensor of the field 
\begin{equation}\label{4.1}
T_{\mu\nu} [\phi]= \partial_\mu \phi\partial_\nu \phi - \frac{1}{2} \eta_{\mu\nu} (\partial \phi)^2~~
\end{equation}
above and below the wave front. Since the normal derivatives are not continuous, $\tau _{\mu\nu}$ is non-trivial.
One can check that the only non-vanishing components are
\begin{equation}\label{2.32}
\tau _{uu} \simeq 2 \Delta\chi ~\partial_u \phi_+ ~~,~~\tau _{ui} = \Delta\chi ~\partial_i\phi_+~~,~~
\tau _{ij} = 2 \delta_{ij}\Delta\chi~ \partial_v \phi_+~~,
\end{equation}
where $\Delta\chi$ is defined in (\ref{2.26}).
In calculating $\tau _{uu}$ terms $O(f^2)$ have been neglected. In this approximation $\partial_u\phi\simeq \partial_u\phi_\pm$. Therefore, $l^\mu \tau _{\mu\nu}=0$, that is $\tau _{\mu\nu}$ is in the tangent space to 
$\mathcal{N}$.  It has the same structure as the surface stress-energy tensor (\ref{1.19a}), 
\begin{equation}\label{2.33}	
\tau _{\mu\nu}=\tilde{\sigma} l_\mu l_\nu+\tilde{j}_i ( l_\mu e_\nu^i+ l_\nu e_\mu^i)+\tilde{p}\delta_{ij}e_\mu^ie_\nu^j~~.
\end{equation}
Vectors $l,e_i$ are introduced in Section \ref{S1.2}, and
\begin{equation}\label{2.34}	
\tilde{\sigma}=\tau _{uu}~~,~~\tilde{j}_i =\tau _{ui}~~,~~\tilde{p}=\frac{1}{2} \delta^{ij}\tau _{ij}~~.
\end{equation}
Quantities (\ref{2.34}) measure abrupt behavior of the scalar energy, current and pressure on the shockwave front.
Thus, for example, when the profile function $f$ does not depend on time, Eqs. (\ref{2.30}) and (\ref{2.32}) yield
the following relations:
\begin{equation}\label{2.35}	
\tilde{\sigma}= 4\pi G  ~ \partial_u(\phi_+^2)~\sigma~~,~~
\tilde{j}_i =-2\pi G ~\partial_i(\phi_+^2)~\sigma~~,~~\tilde{p}= 4\pi G  ~\partial_v (\phi_+^2)~\sigma~~.
\end{equation}
The difference between scalar surface tensor (\ref{2.33})  and ``gravitational'' stress-tensor (\ref{1.19a}) is that
$\tau _{\mu\nu}$ is associated to $\theta$-function-like rather than $\delta$-function-like singularities.

\section{Perturbations of fields of point-like sources} \label{S3}
\setcounter{equation}0
\subsection{Shockwaves with elementary profiles} \label{S3.1}

We consider a point-like scalar ``charge'' $Q$, which is at rest at the center of coordinates, with Coulomb  potential  $\phi(x)$ defined in (\ref{2.17}),
and a plane shock gravitational wave which disturbs the field $\phi(x)$. The aim in this Section is to develop a calculation technique for perturbations $\phi_{sw}$ of the potential caused by different shockwaves and find their late-time asymptotics. We follow convention of 
Section \ref{CD} and use the frame of reference where the charge is at 
rest after crossing the wave front.

Note that any profile function $f(\mathbf{x})$ can be written as the superposition of ``elementary profiles'' $f_p({\bf x})$,
\begin{equation} \label{3.1}
f(\mathbf{x}) = \frac{1}{(2\pi)^3} \int d \mathbf{p}~ f_p({\bf x}) \tilde{f}(\mathbf{p})~~, 
\end{equation}
\begin{equation} \label{3.1a}
f_p({\bf x})\equiv e^{i{\bf x}{\bf p}}~~,
\end{equation}
where $d \mathbf{p}=dp_v dp_1 dp_2$.
Technically it is more convenient to find perturbations caused by a shockwave with an (abstract) profile function $f_p({\bf x})$, and construct from them a solution for the required profile. We denote  $\phi_{sw}(x,\mathbf{p})$ perturbations  of the potential $\phi$ for $f_p({\bf x})$. Their Cauchy problem is
\begin{equation}\label{3.3}
\Box \phi_{sw}(x,\mathbf{p})= 0~~,~~
\end{equation}
\begin{equation}\label{3.3a}
\phi_{sw}(x,\mathbf{p})=(f_p({\bf x}) -1-ip_a x^a)\partial_v\phi(x)~~,~~u=0~~,
\end{equation}
see (\ref{2.18}), (\ref{2.21}). 
Consider a unit sphere, $n_v^2+n_y^2+n_z^2=1$, with $n_v =\sin\theta' \cos\phi'$, $n_y = \cos\theta'$, $n_z = \sin\theta'\sin\phi'$ and
a null 4-vector ${\bf l}={\bf l}(n)$ with components
\begin{equation}\label{3.9}
l_u= n_i^2/2 n_v~~,~~l_v = n_v/2~~,~~l_i = n_i~~,~~{\bf l}^2=0~~.
\end{equation}
Then, as is shown in  Appendix \ref{App1}, the perturbation has the following representation:
\begin{equation} \label{3.14}
	\phi_{sw}(x,\mathbf{p})  =  \int_{S^2} d \Omega'~ 
	\theta(x \cdot {\bf l}) A(\mathbf{p},\Omega') 
	    ~e^{i k_+  (x \cdot {\bf l})} ~~.
\end{equation}
The integral goes over the sphere, $d\Omega'=\sin\theta' d\theta' d\phi'$, and
\begin{equation} \label{3.15}
	A(\mathbf{p},\Omega') 
\equiv  \frac{Q  }{8\pi^2}\frac{ k_{+}^2 (k_{+} n_v-2 p_v)    }{k_+-k_-}~~,
\end{equation}
where
\begin{eqnarray}
	\label{3.8a}
	k_\pm =\breve p_{||}\pm i \breve p_{\perp}=(\breve p \cdot n)\pm i \sqrt{	\breve{p}^2-(\breve p \cdot n)^2}~~,~~\breve p_v =2 p_v~~,~~\breve p_i = p_i~~.
\end{eqnarray}
%
%

Solution \eqref{3.14} has a suitable form to construct perturbations from shockwaves of generic profiles with the help of \eqref{3.1},
\begin{equation}\label{3.2}
\phi_{sw}(x) = \frac{1}{(2\pi)^3} \int d \mathbf{p} ~  \tilde{f}(\mathbf{p}) \phi_{sw}(x,\mathbf{p})~~.
\end{equation}
Before we proceed with analysis of particular shockwaves we discuss one important property 
of \eqref{3.14}.

\subsection{Shocks when shockwave hits the source} \label{S3.2}

\begin{figure} 
	\begin{center}
		\includegraphics[width=6cm]{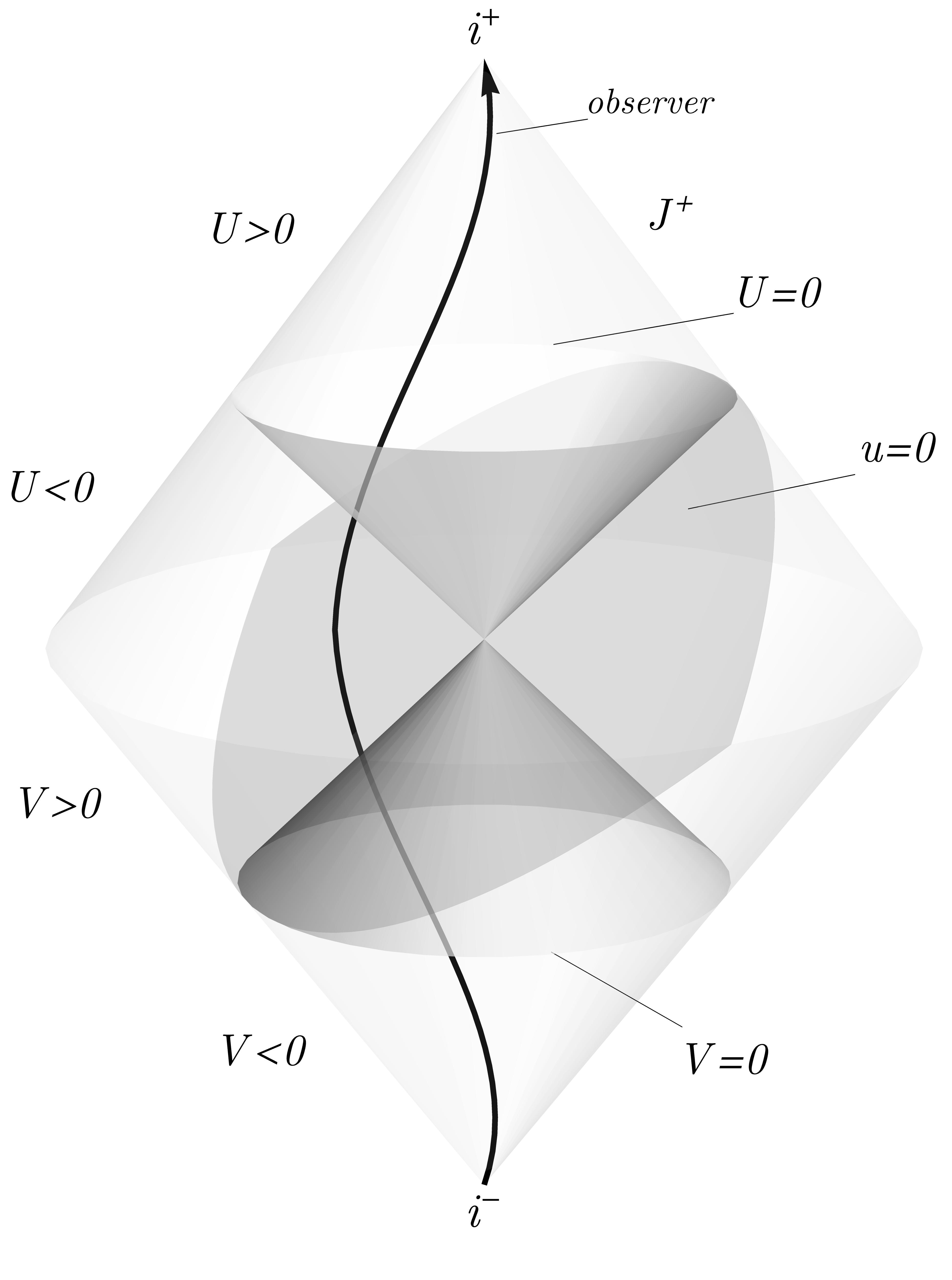}
	\end{center}
	\caption{shows the future, $U=0$, and 
the past, $V=0$, light cones of the event $x=0$, when the source crosses the shockwave front $u=0$. The both cones are tangent to $u=0$. Only the future light cone lies in the domain $u>0$ where perturbation $\phi_{sw}$ is present. The future of any observer crossing the shockwave front lies inside the cone $U>0$. The cone $U=0$ is the front of the second scalar shockwave.} \label{f2}
\end{figure}

When the gravitational shockwave hits the point-like source one may expect new physical effects. Technically these effects are 
related to non-smooth behavior
of the perturbation on a future null cone due to the presence of the $\theta$-function in integral \eqref{3.14}.   Let 
$r=\sqrt{x^2+y_i^2}$ be a distance from the source to some point with coordinates $x,y^i$.
It is convenient to introduce two unit vectors, 
\begin{equation}\label{3.17}
\vec{e}=\left(e_x={x \over r},e_i={y^i \over r} \right)~~,~~\vec{m}=(m_x, m_i)~~,~~m_x = n^2_v-n_i^2~~,~~ m_i=2 n_v n_i~~,
\end{equation}
$\vec{e}^{\phantom{.}2}=\vec{m}^2=1$, where $n_a$ are defined before Eq. (\ref{3.9}). Direction of $\vec{m}=\vec{m}(\Omega')$ varies
when one integrates over $S^2$ in \eqref{3.14}. Vector $\vec{e}=\vec{e}(\Omega)$ defines direction
from the source to the point.  With these notations and definitions \eqref{3.9} the argument of the $\theta$-function in \eqref{3.14}
takes the form:
\begin{equation}\label{3.40}
(x\cdot  {\bf l}) = \frac{1}{2 n_v}\Bigl(U+r \bigl(1+(\vec{m}\cdot \vec{e}) \bigr)\Bigr)=
\frac{1}{2 n_v}\Bigl(V-r \bigl(1-(\vec{m}\cdot \vec{e}) \bigr)\Bigr)~~,
\end{equation}
where $U=t-r$, $V=t+r$ are retarded and advanced coordinates. The null hypersurfaces $U=0$ and $V=0$ are, respectively, the future and 
the past light cones of the event $x=0$ when the wave hits the source. The regions $U>0$ and $V<0$ lie inside the light cones, see Fig. \ref{f2}. Only the future cone is of our interest since it lies at $u>0$.

It is important  that $\theta(x\cdot  {\bf l})\neq 0$ if  $U>0$, $V<0$. Consider the derivative of the  perturbation 
\begin{equation} \label{3.41}
	\partial_U\phi_{sw}(x,\mathbf{p})  =  \partial_U \phi^{(1)}_{sw}+\partial_U \phi^{(2)}_{sw}~~,
\end{equation}
\begin{align}
	\partial_U \phi^{(1)}_{sw} (x,\mathbf{p}) &=  \int_{S^2} d \Omega'~ 
	\theta(x \cdot {\bf l}) \, {iA k_+ \over 2n_v}
	~e^{i k_+  (x \cdot {\bf l})}~~, \label{3.42}
	\\
	\partial_U \phi^{(2)}_{sw} (x,\mathbf{p}) &=  \int_{S^2} d \Omega'~ 
	\delta(x \cdot {\bf l}) \, \frac{A }{2n_v} ~~. \label{3.43}
\end{align}
(The poles in (\ref{3.42}) at $n_v=0$ do not affect convergence of the integral because 
of the exponential factor.)  Part $\partial_U \phi^{(1)}_{sw}$ is continuous on the cone $U=0$. As for (\ref{3.43}) it follows from (\ref{3.40})
that  
\begin{equation} \label{3.44}
\lim_{U\to 0 +} \partial_U \phi^{(2)}_{sw}(x,\mathbf{p}) =0~~,
\end{equation}
\begin{equation} \label{3.45}
\lim_{U\to 0 -} \partial_U \phi^{(2)}_{sw} (x,\mathbf{p}) =- {\Phi_s(\Omega,\mathbf{p}) \over r}~~,~~
\Phi_s(\Omega,\mathbf{p})\equiv -\int_{S^2} d \Omega'~ 
\delta(1+(\vec{m}\cdot \vec{e})) \sign (n_v) A~~.
\end{equation}
Therefore, the normal derivative has the following jump  on the cone:
\begin{equation} \label{3.46}
[\partial_U \phi_{sw}(x,\mathbf{p}) ]\equiv \lim_{U\to 0 +} \partial_U \phi_{sw} (x,\mathbf{p}) -\lim_{U\to 0 -} \partial_U \phi_{sw} 
(x,\mathbf{p})= {\Phi_s(\Omega,\mathbf{p}) \over r}~~.
\end{equation}
When the gravitational shockwave hits the source it creates the second scalar shockwave with the wave front on the cone $U=0$.
For shockwaves with an arbitrary profile function the jump of normal derivatives is determined with the help of (\ref{3.2}),
\begin{equation}\label{3.46b}
[\partial_U \phi_{sw}(x) ] = \frac{1}{(2\pi)^3} \int d \mathbf{p} ~  \tilde{f}(\mathbf{p})~ [\partial_U \phi_{sw}(x,\mathbf{p}) ]=
{\Phi_s(\Omega) \over r}~~,
\end{equation}
\begin{equation}\label{3.46h}
\Phi_s(\Omega)  =-\frac{2}{(2\pi)^3}\int_{S^2} d \Omega'\int d \mathbf{p} ~ 
\delta(1+(\vec{m}\cdot \vec{e}))~\theta(n_v)\tilde{f}(\mathbf{p})~ \Re A(\mathbf{p},\Omega') ~~,
\end{equation}
where we used the property $\vec{m}(n)=\vec{m}(-n)$, see (\ref{3.17}), and the fact that $A(\mathbf{p},-n)=-A^*(\mathbf{p},n)$.
Examples of (\ref{3.46b}) are considered below.

A direct integration in (\ref{3.45}) yields
\begin{equation}\label{3.46d}
\Phi_s(\Omega,\mathbf{p})  =-{\pi \over \tilde{n}_v} \Re
A(\mathbf{p},\Omega')\bigl|_{\vec{m}=-\vec{e}}~~.
\end{equation}
Here one should substitute into $	\Re A(\mathbf{p},\Omega')$  defined by \eqref{3.15} the values of $n_v$ and $n_i$ which satisfy ${\vec{m}( n)=-\vec{e}}$,
\begin{equation}\label{3.46c}
\tilde n_v=\sqrt{1 -(\vec{e} \cdot \vec{v}) \over 2}~~,~~\tilde n_i=-{e_i \over \sqrt{2(1 -(\vec{e} \cdot \vec{v}))}}~~,
\end{equation}
where $\vec{v}$ is the coordinate velocity of the shockwave, $v_x=1,v_i=0$.

As have been discussed in Section \ref{ND} such shocks are characterized by abrupt behavior of the 
components of the stress-energy tensor. By analogy with (\ref{2.31}) one can introduce a tensor structure 
on $U=0$
\begin{equation}\label{3.46f}
\bar{\tau}_{\mu\nu} \equiv \lim_{U\to 0 +} T_{\mu\nu} [\phi_{sw}]- \lim_{U\to 0 -}T_{\mu\nu} [\phi_{sw}]~~,
\end{equation}
where $T_{\mu\nu}$ is given by (\ref{4.1}).  In retarded time coordinates, 
\begin{equation}\label{3.19}
	ds^2 
	=-dU^2 - 2 dU dr+ r^2\gamma_{AB}dx^Adx^B~~,
\end{equation}
the only non-vanishing components of $\bar{\tau} _{\mu\nu}$ are
\begin{equation}\label{3.47}
\bar{\tau} _{UU} \simeq [\partial_U \phi_{sw}](2 \partial_U -\partial_r)\phi_{sw}~~,~~
\bar{\tau} _{UA} =  [\partial_U \phi_{sw}]\partial_A \phi_{sw}~~,~~
\bar{\tau}_{AB} =  r^2\gamma_{AB}[\partial_U \phi_{sw}]\partial_r \phi_{sw}~~,
\end{equation}
where $\gamma_{AB}$ is the metric on unit sphere, $[\partial_U \phi_{sw}]$ are defined with the help of (\ref{3.46b}), and $\partial_\mu \phi_{sw}$ are taken at $U=0$. 
If $\bar{l}_\mu= \delta_\mu^U$ is a null normal to the light cone then one can check that $\bar{l}^\mu \bar{\tau}_{\mu\nu}=0$.
Therefore $\bar{\tau} _{\mu\nu}$ is a tensor in the tangent space to the null hypersurface $U=0$ and it has the same structure as the surface stress-energy tensor (\ref{1.19a}), 
\begin{equation}\label{3.48}	
\bar{\tau} _{\mu\nu}=\bar{\sigma} \bar{l}_\mu \bar{l}_\nu+\bar{j}_A ( \bar{l}_\mu \bar{e}_\nu^A+ \bar{l}_\nu  \bar{e}_\mu^A)+
\bar{p}~\delta_{AB} \bar{e}_\mu^A  \bar{e}_\nu^B~~,
\end{equation}
\begin{equation}\label{3.49}	
\bar{\sigma}=\bar{\tau} _{UU}~~,~~\bar{j}_A =\frac{1}{r}~\bar{\tau} _{UA}~~,~~\bar{p}=\frac{1}{2 r^2}~ \gamma^{AB}\bar{\tau} _{AB}~~,
\end{equation}
where vectors $\bar{l}, \bar{e}^A$ make a basis in the tangent space of the light cone, 
$(\bar{e}^A\cdot \bar{e}^B)=\delta_{AB}$, $(\bar{e}^A\cdot \bar{l})=0$.

\subsection{Late-time asymptotics of perturbations from elementary \\ shockwaves} \label{S3.3}

By using the retarded time coordinates (\ref{3.14}) we estimate large $r$ (or late-time) asymptotics of perturbations \eqref{3.14}.
Large $r$ is the limit when the future null infinity $\mathcal{J}^+$ is approached, see Fig. \ref{f2}.
The technique to derive the asymptotic form of perturbations
generated by null cosmic strings can be found in Appendix A of \cite{Fursaev:2023lxq}. 
It is based on a saddle-point method and can be applied to the case 
of generic gravitational shockwaves. Here we use the same approach.

Due to the exponent in integral \eqref{3.14} the main contribution to the integral comes from a domain where the factor $\vec{m} \cdot \vec{e}+1$ is small, that is, $\vec{m}$ is almost $-\vec{e}$. It is this domain where the saddle-point approximation can be safely used.
Keeping this in mind, we split the integration in \eqref{3.14} into the region $\vec{m} \cdot \vec{e}+1 \leq \Lambda^2$ , and to the rest region, $\vec{m} \cdot \vec{e}+1 > \Lambda^2$, with $\Lambda$ being
a dimensionless parameter in the interval
\begin{equation}\label{3.21}
	\frac{|U|}{r} \ll \Lambda^2 \ll 1~~.
\end{equation}
After some calculations, analogous to those of Appendix A in \cite{Fursaev:2023lxq}, one gets the following estimate for the perturbations \eqref{3.14} :
\begin{equation}\label{3.22}
	\phi_{sw}(x,\mathbf{p}) \simeq \frac{1}{r} \Bigl(	\Phi(U,\Omega,\mathbf{p} ) + \hat{\Phi}(\Omega,\mathbf{p},\Lambda)\Bigr)  + O(r^{-2})~~.
\end{equation}
The  amplitudes $\Phi(U,\Omega,\mathbf{p}), \hat{\Phi}(\Omega,\mathbf{p},\Lambda)$ are determined by the integration 
in the domain $\vec{m} \cdot \vec{e}+1 \leq \Lambda^2$. Contribution to (\ref{3.22}) from the region $\vec{m} \cdot \vec{e}+1 > \Lambda^2$ is exponentially small at large $r$. To write the amplitudes we introduce the unit vector $\tilde{\mathbf{n}}$ \eqref{3.46c}.
One can check with the help of (\ref{3.17}) that $\vec{m}(\tilde{\mathbf{n}})=- \vec{e}$.

Then the amplitudes in (\ref{3.22}) can be written as:
\begin{equation}\label{3.23}
	\Phi(U,\Omega,\mathbf{p} ) = 
	\begin{cases}
		\Phi_+ (U,\Omega,\mathbf{p} )~~,~~U>0 \\
		\Phi_-(U,\Omega,\mathbf{p} ) ~~,~~U<0
	\end{cases}	~~,
\end{equation}
\begin{equation}\label{3.23b}
\Phi_+ (U,\Omega,\mathbf{p} )= 
	\frac{i\pi  \tilde{A} }{\tilde{k}_+}  
	~ e^{i \tilde{k}_+  U /(2 \tilde{n}_v) }~,~
	\Phi_- (U,\Omega,\mathbf{p} )= -
	\frac{i\pi  \tilde{A}^* }{\tilde{k}_-}  
	~ e^{i \tilde{k}_-  U /(2 \tilde{n}_v) }+2\pi i ~\Re \left(\frac{\tilde{A} }{\tilde{k}_+}\right) ~,
\end{equation} 
\begin{equation}\label{3.25}
	\hat{\Phi}(\Omega,\mathbf{p},\Lambda)=	-  
	\frac{ i\pi\tilde{A} }{\tilde{k}_+}   
	~  	
	e^{i \tilde{k}_+  r \Lambda^2 /(2 \tilde{n}_v)}~~.
\end{equation}
Here  $\tilde{A}$,  $\tilde{k}_\pm$ are the values of $A$  and $k_\pm$ at
$\mathbf{n}=\tilde{\mathbf{n}}$, see definitions \eqref{3.15}, \eqref{3.8a}. Although $\hat{\Phi}$ is exponentially small, we see 
in Section \ref{S3.4} that for shockwaves with certain profiles, after integration over $\mathbf{p}$, it may produce a finite static part.

Since the perturbation is continuous at $U$, so does function $\Phi(U,\Omega,\mathbf{p} )$. One can check by using (\ref{3.23b})
that $\Phi_+(0,\Omega,\mathbf{p} )=\Phi_-(0,\Omega,\mathbf{p} )$.
However $\partial_U\Phi(U,\Omega,\mathbf{p} )$ is not continuous at $U=0$. The jump of the derivative, which follows from
(\ref{3.23}),(\ref{3.23b}), is in accord with  (\ref{3.46b})-(\ref{3.46d}).

Asymptotic \eqref{3.22} is one of our main results. 
It describes an outgoing spherical wave produced by a shockwave perturbation of the static potential of a point-like source. 
Amplitude \eqref{3.23} is 
the dynamical part which, as we show, determines a finite outgoing energy flux. 
The cutoff parameter $\Lambda$, present in the static part of (\ref{3.25}), does not affect this flux.

\subsection{No shocks from null strings} \label{S3.4}

It is instructive to use \eqref{3.22} to find perturbation caused by a straight null cosmic string and compare it with results
of \cite{Fursaev:2022ayo,Fursaev:2023lxq,Fursaev:2023oep}. A gravitational field of null string stretched along $y_2=z$ axis, which moves along the $x$ axis with an impact parameter $a$ along the $y_1=y$ axis, is described by a shockwave metric with the profile listed in Table \eqref{t1}. The Fourier transform of the profile function is
\begin{equation}\label{3.34}
	\tilde{f}(\mathbf{p}) =- 8\pi^2 \omega_s \frac{  e^{i p_y a}}{p_y^2} \delta(p_v) \delta(p_z)~~,
\end{equation}
where $\omega_s=8\pi G E_s$ and $E_s$ is the energy per unit length of the null string. 

Although the null string space-time looks as a shockwave geometry the string does not produce any shockwave. The space-time  around the 
null string world-sheet is locally Minkowskian but there is a non-trivial holonomy related to a group of null rotations around the string.
Therefore one does not expect scalar shockwaves when the source crosses the hypersurface $\cal N$. This is indeed the case, since
the following integral:
\begin{equation}\label{3.34b}
\int d \mathbf{p} ~\tilde{f}(\mathbf{p})~ \Re A(\mathbf{p},\Omega') \sim
\delta(a) =0~~,
\end{equation}
which appears in the right hand side of (\ref{3.46h}), vanishes for the string profile function \eqref{3.34}. In this case there is no jump 
of normal derivatives of the perturbation on the light cone, $[\partial_U \phi_{sw}(x) ] =0$.

By using \eqref{3.2} and \eqref{3.22} we get 
\begin{equation} \label{3.35}
	\phi_{sw}^{(s)}(x) \simeq \Re 
	\left\{
	 \frac{ N(\Omega) }{r} 
	\left(
	-\ln \left( \frac{ U- e_y a   + ia  \varepsilon}{a} \right)
	+\ln (r/ \varrho)
	+ s(\Omega,\Lambda)
	\right)
	\right\}  + O( r^{-2})~~,
\end{equation}
\begin{equation}\label{3.36}
	N(\Omega) =  \frac{ \omega_s Q  }{8 \pi^2}  
	\frac{ (1-e_x)^2}{\varepsilon~ (\varepsilon+i e_y)^2 }~~,~~
	\varepsilon = \sqrt{2(1-e_x)-e_y^2}~~,
\end{equation}
where $\varrho=a/\Lambda^2$.  A static part $s(\Omega,\Lambda)$, whose precise form is not important for the subsequent analysis, appears from exponentially small terms, which have been ignored in \eqref{3.22}.
Asymptotic (\ref{3.35}) does coincide with the result of \cite{Fursaev:2023oep}.

The logarithmic term $\ln (r/ \varrho)$ in \eqref{3.35} appears since standard radiation conditions are violated in the presence of null strings, see \cite{Chrusciel:1993hx}. We return to discussion of logarithmic terms for other relativistic  
sources in Section \ref{S5}.

In Section \ref{ND} we also discussed shockwaves in field systems which are caused by discontinuities of normal derivatives at the null surface $u=0$. The absence of such discontinuities in case of null strings has been demonstrated in  \cite{Fursaev:2022ayo,Fursaev:2023lxq,Fursaev:2023oep}.

\section{Physical effects from shockwaves produced by null particles and null branes} \label{S4}
\setcounter{equation}0
\subsection{Definitions}

We use results of the previous Section to study perturbations caused by shockwaves from null particles and null branes. 
We show that perturbations $\phi_{sw}(r,U,\Omega)$ of fields of point scalar charges have form (\ref{i.1}).  The total radiation energy flux through a sphere of large radius
with the center at the position of the charge is determined by the amplitude $\Phi(U,\Omega)$ in (\ref{i.1}),
\begin{equation}\label{4.2}
W(U,R)= \int_{r=R} d\Omega~ r^2~ T^r_U(r,U,\Omega)= \int_{r=R} d\Omega~ (\partial_U\Phi)^2~~.
\end{equation}
This result follows from (\ref{i.1}) and the definition of  the stress-energy tensor at $u>0$ given by (\ref{4.1}).
 Of some use are also an averaged intensity of the flux
\begin{equation}\label{4.3}
	I(\Omega) = \frac{1}{a} \int_{-\infty}^\infty dU~r^2~ T^r_U(r,U,\Omega)~~,
\end{equation}
and the total emitted energy
\begin{equation}\label{4.4}
	E \equiv \int_{-\infty}^\infty dU~ W(U,R) = a \int d\Omega~ I(\Omega)~~.
\end{equation}
Other characteristics of the radiation can be introduced if one considers more realistic models, in case of electromagnetic radiation from electric charges, for example. We leave this analysis for future studies.

\subsection{The case of null particles}\label{S4.1}

We consider a massless particle which moves along the $x$ axis with an impact parameter $a$ in one of the orthogonal directions.
The scalar charge, as earlier, is supposed to be at rest at the center of coordinates after crossing the wave front $\cal N$. The Fourier transform of the shockwave profile in this case is
\begin{equation}\label{3.29}
	\tilde{f}(\mathbf{p}) = -(2\pi)^2 \omega_p \frac{  e^{i p_y a}}{p_i^2} \delta(p_v)~~,
\end{equation}
where $\omega_p\equiv 8\pi G E_p$. The delta-function in the r.h.s of \eqref{3.29} appears because the profile function for a null particle 
does not depend on time.
By using \eqref{3.2} and \eqref{3.22}, we obtain perturbation $\phi_{sw}(x)$ given by  (\ref{i.1}) with the amplitude
\begin{equation}\label{3.30}
\Phi(U,\Omega) = -\frac{\omega_p}{2\pi } \int d^2 p~ \Phi(U,\Omega,\mathbf{p})~ \frac{e^{i p_y a}}{p_i^2}~~.
\end{equation}
By performing the integrating over $p_i$ one finds:
\begin{equation}\label{3.32}
	\Phi(U,\Omega) = 
	\begin{cases}
		\Phi_+ (U,\Omega)~~,~~U>0 \\
		\Phi_- (U,\Omega)~~,~~U<0
	\end{cases}~~,~~
		\Phi_- (U,\Omega)= 	-	\Phi^*_+(U,\Omega) - \frac{\omega_p Q}{8 \pi a} e_y~~,
\end{equation}
\begin{equation}\label{3.32b}
	\Phi_+(U,\Omega)=   
	\frac{\omega_p Q (1-e_x)   }{16\pi }~ \Re
	\Biggl(
	\frac{e_y+i e_z}{ U(e_y+ie_z) - a (1-e_x)  }
	+ 	
	\frac{1}{ c     }	    
	\left(
	1- \frac{2a b}{U} \frac{b e_y-e_z}{1+b^2} 
	\right)	     
	\Biggr)~~,
\end{equation}
\begin{equation}\label{4.9}
b =  -\frac{U   }{a e_z+i c} ~~,~~ 
	c^2 = (U-e_y a)^2+ a^2 (1-e_x)^2 ~~.
\end{equation}
As expected, the amplitude is continuous on the light cone where it has the value:
\begin{equation}
	\Phi_+(0,\Omega) =\Phi_-(0,\Omega) = \frac{\omega_p Q}{16\pi a} 
	\left(
	  \frac{1-e_x-e_z^2}{\sqrt{(1-e_x)^2+e_y^2}}-e_y
	\right)~~.
\end{equation}
To calculate the flux
\eqref{4.2} we need time derivatives of the amplitude. Straightforward but tedious computations yield 
\begin{equation}\label{4.6}
	\partial_U \Phi_+(U,\Omega) = \frac{\omega_p Q (1-e_x)   }{16\pi } ~\Re~ F(U,\Omega)~~,
\end{equation}
\begin{multline} \label{4.8}
	F(U,\Omega) =     
	-\frac{(e_y+i e_z)^2}{  \bigl(U(e_y+i e_z) -a(1-e_x)   \bigr)^2}
	-
	\frac{U-e_y a}{c^3 }	    
	\left(
	- \frac{2a b}{U} \frac{b e_y-e_z}{1+b^2} 
	\right)
	\\
	-
	\frac{1}{c }	    
	\left(
	- \frac{2a} {U^2} \frac{b(b e_y-e_z)}{1+b^2} 
	 +
	 \frac{2a }{U}  \partial_U b \frac{b^2 e_z+2 b e_y-e_z}{(1+b^2)^2} 
	\right)~~,
\end{multline}
and $\partial_U\Phi_- (U,\Omega)= 	-	\partial_U\Phi^*_+(U,\Omega) $. At the light cone one finds
\begin{equation} \label{4.6b}
	\partial_U \Phi_+(0,\Omega) = \frac{\omega_p Q   }{16\pi a^2 (1-e_x)}
	\left(
	  e_z^2-e_y^2 + e_y \frac{3(1-e_x-e_z^2)^2-e_z^4}{((1-e_x)^2+e_y^2)^{3/2}}
	\right) ~~,
\end{equation}
\begin{equation} \label{4.6c}
	\partial_U \Phi_-(0,\Omega) = \partial_U \Phi_+(0,\Omega) -[\partial_U \Phi](\Omega)~~,
\end{equation}
\begin{equation} \label{4.6d}
[\partial_U \Phi](\Omega)=
	\frac{ \omega_p Q  }{8 \pi a^2  }
	\frac{e_z^2-e_y^2}{1-e_x}
	~~.
\end{equation}
Thus the jump of the normal derivative is non-trivial, which means that the shockwave from a massless particle produce 
a secondary scalar shockwave when it hits the source, see Section \ref{S3.2}.

\begin{figure} 
	\begin{center}
		\includegraphics[width=6cm]{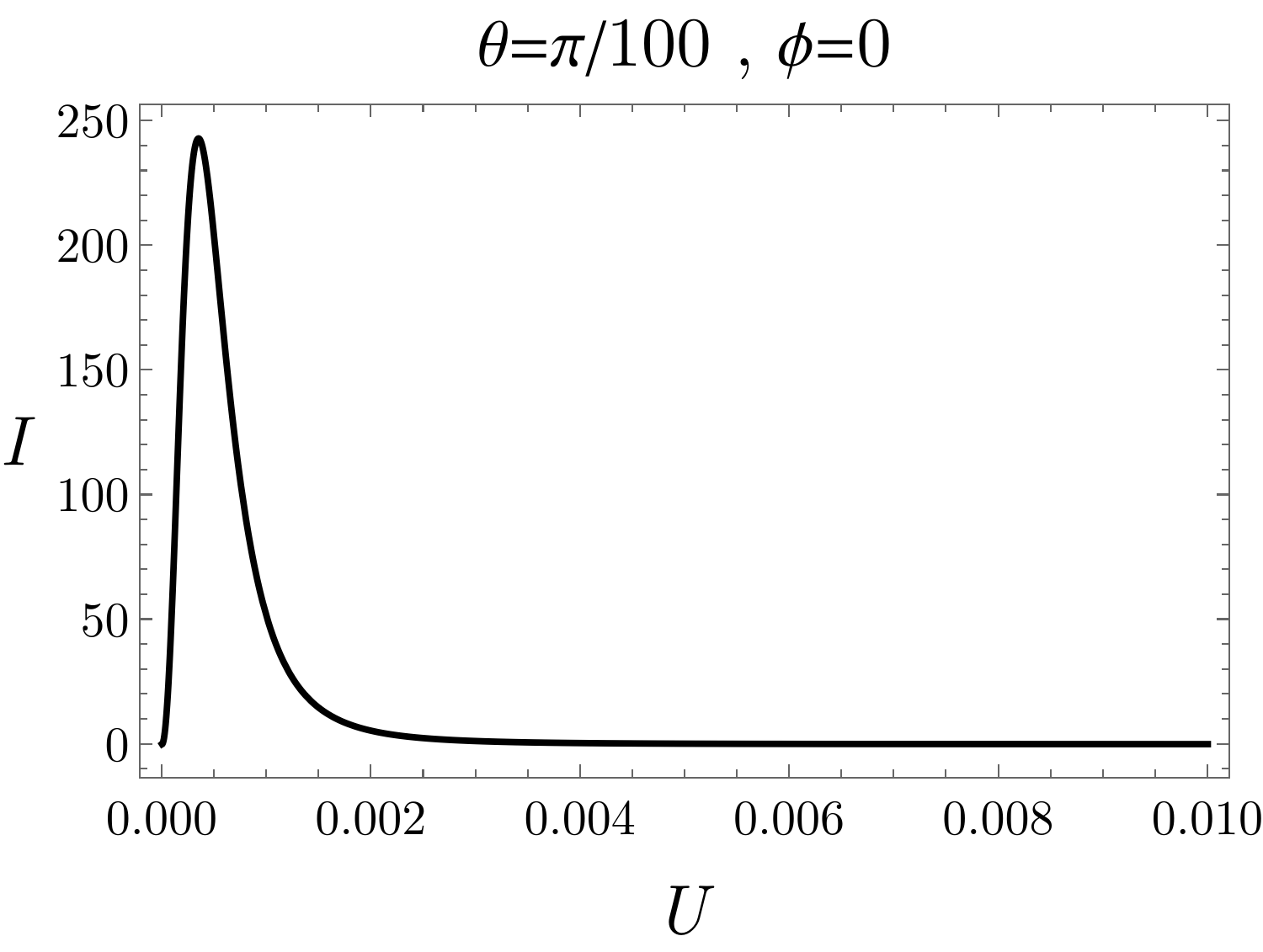}\hspace{1cm}
		\includegraphics[width=6.3cm]{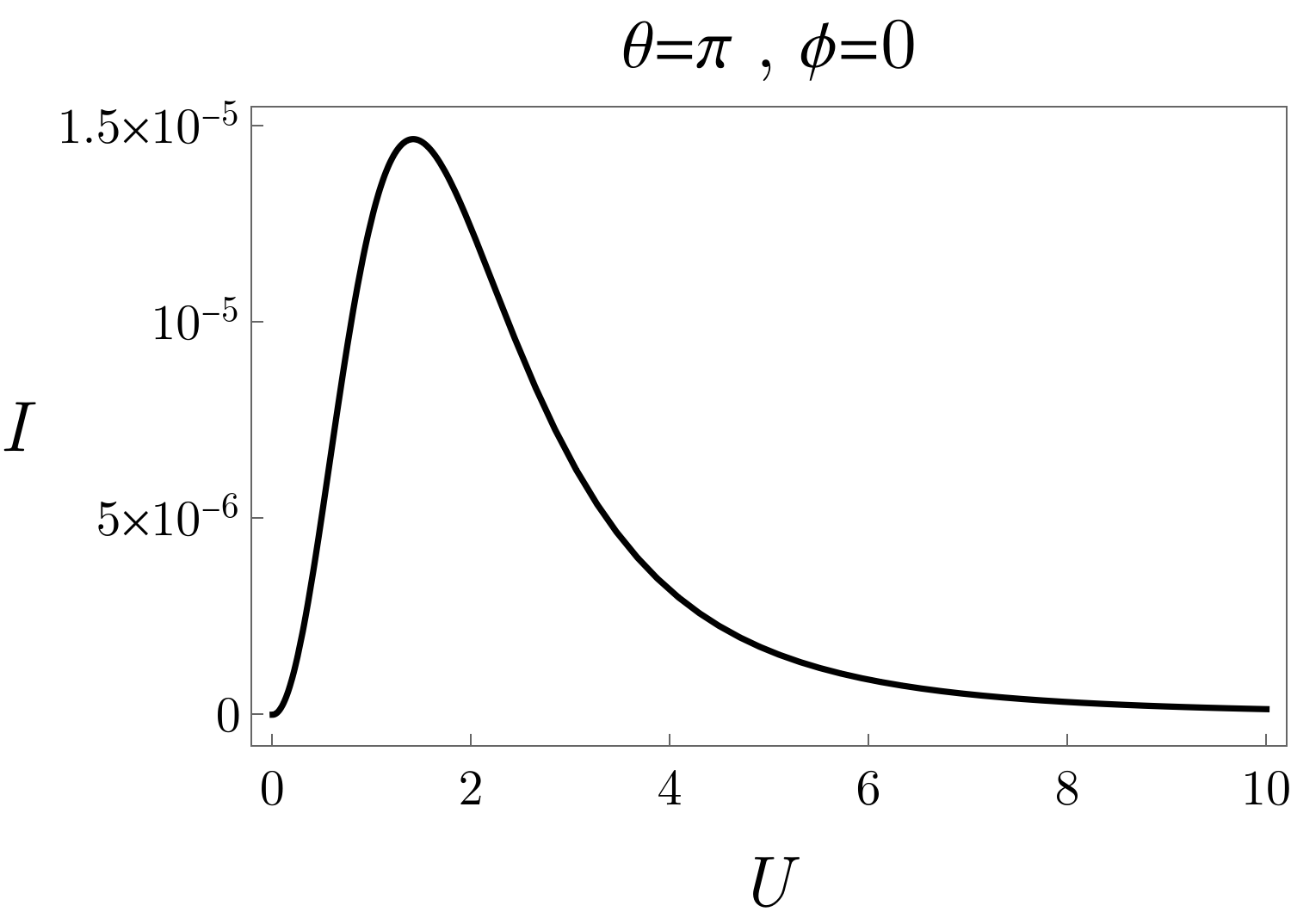}
	\end{center}
	\caption{shows the intensity of the flux (\ref{4.2p}) inside the future light cone, $U>0$. The left part is the intensity in the direction of the null particle, $e_x\to 1$, the right part is the intensity in the opposite direction,  $e_x\to - 1$. The intensity in the direction of particle motion is several orders of magnitude higher.} \label{f1}
\end{figure}

The intensity of the radiation  energy flux defined in (\ref{4.2}) is
\begin{equation}\label{4.2p}
I(U,\Omega)=\frac{\omega^2_p Q^2   }{256\pi^2 }~ (1-e_x)^2(\Re~ F(U,\Omega))^2~~.
\end{equation}
At large $U$ it decays as $O(U^{-4})$.
One can show that the flux is concentrated in the direction of motion of the null particle which creates the shock gravitational wave,
see Fig. \ref{f1}.
The property that the radiation is mostly directed toward the motion of the relativistic source also holds in case of the radiation from 
null strings.

\subsection{The case of null branes}\label{S4.2}

Another important example of relativistic source producing shockwaves are null branes (ultrarelativistic domain walls).
The corresponding profile function  has the form of second order polynomial
\begin{equation}\label{f_br}
f({\bf x})=f_{ab}x^a x^b~~. 
\end{equation}
If $f_{vb}=0$ and the rest coefficients are constant, the corresponding null brane has constant energy density $\sigma$ and vanishing 
surface currents and pressure, $j_i=p=0$.

One can consider examples of null branes with nontrivial currents and pressure. One of the options is to allow nonvanishing constants
$f_{vb}$. This results in the following parameters, according to \eqref{1.19b}  :
\begin{equation}\label{5.13}
\sigma=\frac{f_{ii}}{8\pi G}~~,~~ j_i=-\frac{f_{vi}}{4\pi G}~~,~~
	p=\frac{f_{vv}}{2\pi G}~~,
\end{equation}
\begin{equation}\label{5.14}
f_{ii}f_{vv}-f^2_{vi}\geq 0~~,~~f_{vv}<0~~,
\end{equation}
where \eqref{5.14} is a consequence of \eqref{1.26}.  Another example of shockwaves with nontrivial surface currents and pressure
are waves with time dependent energy parameter $E_b$ in the profile function, see Table \ref{t1},
with restriction $\ddot E_b E_b - \dot E_b^2 \ge 0$ implied by \eqref{1.26}. Other examples are also possible.

As an illustration, consider the perturbation of the field of the source~\eqref{2.17} by a shockwave with a profile function~\eqref{f_br}.
We assume that $f_{ab}$ are some constants, 
use \eqref{3.2} and the solution for the elementary profile \eqref{3.14},  \eqref{3.15}.
In the given case the Fourier transform of the shockwave profile
\eqref{f_br}	is 
\begin{equation}
	\tilde f({\bf p})=- (2\pi)^3 f_{ab}\frac{\partial}{\partial p_a}\frac{\partial}{\partial p_b} \delta^{(3)}({\bf p})~~,~~ a,b= v,y^i~~. 
	\label{fp}
\end{equation}
Substituting \eqref{fp} in \eqref{3.2} one finds
\begin{equation} \label{5.17}
	\phi_{sw} (x) = -\lim_{p_a\to 0} \left(	f_{ab}\frac{\partial}{\partial p_a}\frac{\partial}{\partial p_b}\phi_{sw}(x,\mathbf{p})\right)~~,
\end{equation}
where $\phi_{sw}(x,\mathbf{p})$ is defined by  \eqref{3.14}, \eqref{3.15}.  Leading terms of  $\phi_{sw}(x,\mathbf{p})$ at $p_a\to0$ 
are quite simple. Since  $e^{ik_{\pm}(x\cdot l)}\simeq 1+O(p)$, with making use of the  property $A(\mathbf{p},-n)=-A^*(\mathbf{p},n)$, the  solution \eqref{3.14}  can be rewritten as
\begin{align} \label{5.18}
	\phi_{sw}(x,\mathbf{p})  &=   C(\mathbf{p})+\int_{S^2} d \Omega'~ 
	\theta(x \cdot {\bf l}) ~\Re A(\mathbf{p},\Omega') +O(p^3)~~,\\
	\label{5.20}
	C(\mathbf{p}) &=  i \int_{S^2} d \Omega'~ \Im A(\mathbf{p},\Omega') ~~.
\end{align}	
Thus, one gets for (\ref{5.17})
\begin{equation} \label{5.21aa}
	\phi_{sw} (x) =-\int_{S^2} d \Omega'~ 
	\theta(x \cdot {\bf l}) \left(	f_{ab}\frac{\partial}{\partial p_a}\frac{\partial}{\partial p_b} \Re A(\mathbf{p},\Omega')
	\right)~~.
\end{equation}
Note that a pure imaginary constant contribution to $\phi_{sw} (x)$ from $C(\mathbf{p})$ in (\ref{5.18}) vanishes since 
the perturbation is a real scalar field. The derivative in \eqref{5.21aa} reads
\begin{equation}
	f_{ab}\frac{\partial}{\partial p_a}\frac{\partial}{\partial p_b} \Re A(\mathbf{p},\Omega')
	=
	\frac{Q}{16\pi^2}
	\Bigl(
	\breve n_v \bigl(4 f_{ab}\breve n_a \breve n_b-4 f_{vv}-f_{ij}\delta^{ij} \bigr)-8 f_{av}\breve n_a
	\Bigr)~~,
	\label{5.21b}
\end{equation}
where $\breve{n}_v=2n_v~,~\breve{n}_i=n_i$.  

We now go to  retarded time coordinates (\ref{3.19}). It follows from (\ref{3.40}) that $\theta(x \cdot {\bf l})=\theta(n_v)$ if $U>0$.
Therefore, the perturbation  in the future light cone is
\begin{equation} \label{5.22bb}
	\phi_{sw} (x)=  \frac{Q}{16\pi^2}\int\limits_0^{2\pi} d \phi' \int\limits_0^{\pi/2} 
	d \theta' \sin \theta' ~
	\Bigl(
	8 f_{av}\breve n_a-\breve n_v \bigl(4 f_{ab}\breve n_a \breve n_b-4 f_{vv}-f_{ij}\delta^{ij} \bigr)
	\Bigr) 
	~~,
\end{equation}
and by performing the integration one can show that it completely vanishes,
\begin{equation} \label{5.22bbb}
	\phi_{sw} (x)=0~~,~~U>0~~.
\end{equation}
Outside the light cone the perturbation is non-trivial and depends on coordinates. There is a non-vanishing flux of the radiation 
determined by the  derivative 
\begin{equation} \label{5.22c}
	\partial_U\phi_{sw} (x) = -f_{ab}\int_{S^2} d \Omega'~ 
	\sign(n_v) ~\delta\Bigl(U+r \bigl(1+(\vec{m}\cdot \vec{e}) \bigr)\Bigr)
	\left(\frac{\partial}{\partial p_a}\frac{\partial}{\partial p_b} \Re A(\mathbf{p},\Omega')
	\right)~~.
\end{equation}
To cast (\ref{5.22c}) in an analytic form we change integration over half of the sphere $\vec n^2=1$ , $n_v>0$ 
 to entire sphere $\vec m^2=1 $,  as explained in Appendix \ref{App1}, and get
\begin{multline} \label{5.22cc}
	\partial_U\phi_{sw} (x) 
	= - \frac{Q}{8\pi^2}\int_{S^2} d \Omega''~ 
	\delta\Bigl(U+r \bigl(1+(\vec{m}\cdot \vec{e}) \bigr)\Bigr)\times\\
	\left\{
	\left(-\frac{ f_{ii}}{4}   
	+  (1-4 n_i^2) f_{vv} 
	-  (1-4 n_v^2)\frac{n_i}{n_v} f_{iv}
	\right)  +   f_{ij}n_i n_j  \right\}~~,~~U<0~~.
\end{multline}
 The argument of  delta function in \eqref{5.22cc} vanishes at $\vec m=- \kappa\vec e$. It corresponds to
 \begin{equation}
 	n_v=\sqrt{\frac{1-\kappa e_x}{2}}~~,~~ n_i=-\frac{\kappa e_i}{\sqrt{2(1-\kappa e_x)}}~~,~~\kappa=\frac{U}{r}+1~~.
 \end{equation}
 Then the integration over a unit sphere yields
\begin{multline} \label{5.22d}
	\partial_U\phi_{sw} (x) =
	  \frac{Q}{8\pi  }
	\left\{\frac{ G }{ r} 
	\left( \sigma 
	-2  \left(1- \frac{2\kappa^2 e_i^2}{1-\kappa e_x} \right) p 
	-4 \left( \frac{1-2\kappa e_x}{1-\kappa e_x}\right)\kappa e_i j_i
	\right)  -
	\right.
	\\
	\left.
	-\frac{  f_{ij}}{2\pi r}   \frac{\kappa^2 e_i e_j}{1-\kappa e_x} \right\}~~.
\end{multline}
Here we expressed  the coefficients  in terms  of the parameters \eqref{5.13}.
In the leading order 
\begin{align}
	\partial_U\phi_{sw} (x) &=\frac{\Phi_s(\Omega)}{r}~~,
	\\
		\Phi_s(\Omega) &= \frac{Q}{8\pi  }
	\left\{G 
	\left( \sigma 
	+2  \left(1+ 2 e_x \right) p 
	-4 \left( \frac{1-2 e_x}{1- e_x}\right) e_i j_i
	\right) -\frac{  f_{ij}}{2\pi }   \frac{ e_i e_j}{1- e_x} \right\}~,~U<0~. \nonumber
\end{align}
One can summarize this picture in the following way. When a source crosses a brane it creates a spherical bubble  whose radius grows at the speed of light. The world-sheet of the bubble is the future light cone $U=0$. Outside the bubble,
$U<0$ there is a non-stationary perturbation $\phi_{sw}\neq 0$ of the potential of source which yields a non-vanishing flux of the radiation
to the future null infinity. However, inside the bubble, $U>0$,  the potential of the source is unperturbed, $\phi_{sw}= 0$. 
Any observer after some time inevitably ends up inside the bubble and see unperturbed field of the source.

Interestingly, a similar effect arises when an ultrarelativistic particle is piercing a stationary brane \cite{Galtsov:2013ary}. 
This problem can be viewed as dual to the model presented above, and it is remarkable that the result is similar: an outgoing spherical wave arises due to the excitation of the brane at the moment of its ``perforation'' by the particle.

\section{Discussion} \label{S5}
\setcounter{equation}0

The aim of this work was to suggest a formulation of classical field theory in a background of plane-fronted shock gravitational waves
produced by different ultrarelativistic sources and to describe new physical effects.  One of the effects, a
conversion of non-stationary perturbations into outgoing radiation, has been already observed in case of shockwave geometries 
produced by null cosmic strings. As has been expected, the effect is quite general, and we confirmed this fact by direct computations.
A number of results here have been obtained in an analytical form, see Eqs. (\ref{3.30})--(\ref{3.32b}).

The two new effects, which are absent for null strings, are described. A planar gravitational shockwave creates two shocks in 
the field system. The first shock is a planar scalar shockwave accompanying the initial gravitational wave. The second shock is  a spherical scalar shockwave which appears  when the gravitational wave hits the source. The both shocks are specified by jumps of components of the stress-energy tensor of the field which are tangent to the wave fronts. From geometrical  point of view this corresponds to $\theta$-function-like non-analyticity of the background curvature.

The present work sets a stage for other investigations.
One of the perspectives is to extend the analysis of the scalar field theory to the Maxwell theory and to the linearized Einstein gravity. By taking into account our results 
one may expect that perturbations of electric and gravitational fields generated by shockwaves result 
in outgoing electromagnetic and gravitational waves, similar to what has been demonstrated in \cite{Fursaev:2023oep,Fursaev:2023lxq,Fursaev:2022ayo} for perturbations caused by null strings. Potentially it may find interesting applications.

For example, the problem of scattering of a null particle on a massive object may have a practical significance for astrophysics. Numerical Relativity makes it possible to simulate various complicated processes, such as the formation of jets, the black hole encounters, etc. However, the study of simplified models remains an important tool. In particular, it would be interesting to estimate gravitational radiation from ultrarelativistic sources~\cite{Smarr:1977fy,DEath:1976bbo,Segalis:2001ns,Berti:2010ce} by using our approach.

The presented approach may also be useful for studying the dynamics of bubble walls that can form during the era of cosmological phase transitions. Under certain conditions, bubbles of a new phase can expand almost at the speed of light~\cite{Bodeker:2017cim}, which allows us to consider their walls locally as planar null branes, in some approximation. The bubble wall dynamics affects a number of astrophysical phenomena in the early Universe, such as  particle production, the formation of primordial black holes, the emission of gravitational waves, and others. Therefore, an important task is to study the forces acting on the bubble wall~\cite{Gouttenoire:2021kjv,Long:2024sqg}. Our results indicate a new mechanism that leads to the emergence of outgoing radiation during the interaction of an ultrarelativistic bubble wall with matter fields. This effect can contribute to the energy budget of cosmological first-order phase transitions~\cite{Espinosa:2010hh} for ranaway bubbles and provide additional friction that will be experienced by the bubble wall moving with a large Lorentz factor.
A similar mechanism (called ``piercing gravitational radiation'') was recently discovered \cite{Galtsov:2017udh} in a study of brane perforation by ultrarelativistic particles.

There are also  more mathematical issues. For example, in case of null strings asymptotic form of  perturbation \eqref{3.35} 
includes the logarithmic term $\ln (r/ \varrho)$ which  appears since standard radiation conditions are violated, see \cite{Chrusciel:1993hx}.
The analogous term presents in late-time asymptotics of gravitational perturbations \cite{Fursaev:2023oep,Fursaev:2023lxq}. The corresponding space-time has more complicated structure as the future null infinity is approached. It belongs to a class of polyhomogeneous
space-times. We suggest that asymptotics of other ultrarelativistic sources may include logarithmic terms, depending on
the infrared behaviour of the profile  function $\tilde{f}(\mathbf{p})$ in the momentum representation.
For example, similar logarithmic behaviour at large $r$ appears for shockwave geometries produced by a global monopole \cite{Lousto:1990wn,Barriola:1989hx}, the profile function being $\tilde{f}(\mathbf{p}) = 4\pi^2  \delta(p_v)|p_i|^{-3}$.

We plan to return to these topics in further publications.

\newpage
\appendix

\section{Derivation of perturbations: Eqs. (\ref{3.14}) and (\ref{3.46d})}\label{App1}
\setcounter{equation}0

Here we give the derivation of our basic result (\ref{3.14}). We follow notations of Section \ref{S3.1}.
It is useful to split (\ref{3.3}), (\ref{3.3a}) into two problems
\begin{equation}\label{3.3b}
\phi_{sw}(x,\mathbf{p})=	\Psi_1(x,\mathbf{p}) -\Psi_2(x,\mathbf{p})~~,~~
\Box 	\Psi_{1,2}(x,\mathbf{p})= 0~~,
\end{equation}
\begin{equation}\label{3.3c}
\Psi_1(x,\mathbf{p})=f_p({\bf x})\partial_v\phi(x)~~,~~u=0~~,
\end{equation}
\begin{equation}\label{3.3cd}
\Psi_2(x,\mathbf{p})= (1+ip_a x^a)\partial_v\phi(x)~~,~~u=0~~.
\end{equation}
One of the solutions can be written as, see below,
\begin{equation} \label{3.13}
	\Psi_2(x,\mathbf{p})=	-\frac{ Q }{8\pi^2} \int_{S^2} d \Omega'~ \Bigl(    \partial_v 
	+i   (2 \breve p_{||} l_v -p_v)   \Bigr) \delta(x \cdot \mathbf{l})~~,
\end{equation}
where the integral goes over the sphere, $d\Omega'=\sin\theta' d\theta' d\phi'$, and $\breve p_{||}$ is given by (\ref{3.8a}).
Solution $\Psi_1(x,\mathbf{p})$ can be put in the form:
\begin{equation}\label{3.5}
\Psi_1(x,\mathbf{p}) = \frac{1}{(2\pi)^3} \int d \mathbf{k}~ e^{i (x\cdot k) } ~\tilde{\Psi}_1(\mathbf{k},\mathbf{p})~~,
\end{equation}
where $k_\mu $ is a null vector, $k^2=0$, and 
\begin{equation}\label{3.6}
(x\cdot k)\equiv x^\mu k_\mu  = k_u u + \mathbf{k} \cdot \mathbf{x}~~,~~ k_u =k_i^2/(4 k_v)~~.
\end{equation}
Fourier transform $\tilde{\Psi}_1(\mathbf{k},\mathbf{p}) $ is determined by the initial data \eqref{3.3c}  and by Eq. (\ref{2.17}),
\begin{equation}\label{3.7}
	\tilde{\Psi}_1(\mathbf{k},\mathbf{p}) = \int d \mathbf{x'}~ e^{- i \mathbf{k} \cdot \mathbf{x'}}~ \Psi(0, \mathbf{x'},\mathbf{p}) = \frac{2i Q (k_v-p_v)}{(2(p_v-k_v))^2+|k_i-p_i|^2}~~.
\end{equation}
By writing the integral \eqref{3.5} over momentum space  in spherical coordinates,  $\mathbf{k} = k \mathbf{n}$,  
we have
\begin{equation}\label{3.8}
	\Psi_1(x,\mathbf{p}) 
	= -\frac{iQ}{16\pi^3} \int_{S^2} d \Omega'  \int_{-\infty}^{+\infty} dk~  \frac{ k^2(k l_v - p_v)  }{(k-k_+)(k-k_-)} e^{i k  (x \cdot \mathbf{l})}~~.
\end{equation}
To extend the integration over $k$ in (\ref{3.8}) to the entire axis we used  properties of the integral over $S^2$ under reflection  $\mathbf{n} \to -\mathbf{n}$.  
The integrand in \eqref{3.8} can be further split onto two terms,
\begin{equation}\label{3.11}
	\frac{ k^2(k l_v - p_v)  }{(k-k_+)(k-k_-)} = \frac{ k(k l_v - p_v)   }{k_+-k_-} \left( \frac{ k_+  }{k-k_+} - \frac{ k_-  }{k-k_-}  \right)~~.
\end{equation}
Integrals over $k$ in the each term can be performed by shifting $k \to k+k_\pm$. The answer can be represented as the sum of two terms:
\begin{equation}\label{3.12} 
		\Psi_1(x,\mathbf{p} ) 
		=\Psi_2(x,\mathbf{p})  + 	\phi_{sw}(x,\mathbf{p}) ~~,
\end{equation}
in accord with (\ref{3.3b}), where $\phi_{sw}(x,\mathbf{p})$ is given by (\ref{3.14}).

We now proceed with explanation of formula (\ref{3.13}) for $\Psi_2(x,\mathbf{p})$. The fact that $\Box\Psi_2(x,\mathbf{p})=0$
is obvious since $\bf l$ is null. We need to verify the validity of initial condition (\ref{3.3cd}). So we write
\begin{equation} \label{a1}
	\Psi_2(x,\mathbf{p})=	-\frac{ Q }{8\pi^2} \int_{S^2} d \Omega'~ \Bigl(    \partial_v 
	+i   (8 p_v l_v^2 + 2 p_i l^i l_v -p_v)   \Bigr) \delta(x \cdot {\bf l})~~,
\end{equation}
and use integrals
\begin{align} 
	\int_{S^2} d \Omega' ~	 \delta(x \cdot {\bf l}) &= \frac{2}{\pi}  \int d\mathbf{k} ~\frac{e^{i (x \cdot k)}}{ (2k_v)^2+k_i^2}~~, \label{a2_1}
	\\	
	 \int_{S^2} d \Omega' ~	\partial_a \delta(x \cdot {\bf l}) &= \frac{2}{\pi} \partial_a \int d\mathbf{k} ~
	 \frac{e^{i (x \cdot k)}}{ (2k_v)^2+k_i^2}~~,
	\\
	\int_{S^2} d \Omega' ~	l_v l_a \delta(x \cdot {\bf l}) &= -\frac{2}{\pi} \partial_v \partial_a \int d\mathbf{k} ~
	\frac{e^{i (x \cdot k)}}{\bigl[ (2k_v)^2+k_i^2\bigr]^2}~~, \label{a2_2}
\end{align}
where $k_\mu$ is defined in \eqref{3.6} and $\bf l$ in \eqref{3.9}.  To obtain Eq.\eqref{a1} at the shockwave front, one starts with the Fourier transform of the potential of a point source \eqref{2.17}
\begin{equation}
	\phi(x) =-\frac{Q}{4\pi^3}\int d\mathbf{k} \frac{e^{i (x \cdot k)}}{(2k_v)^2+k_i^2}~~,~~ u=0~~.
\end{equation}
Using this representation and \eqref{a2_1}--\eqref{a2_2}, we come to the following set of integrals at $u=0$:
\begin{align}
		Q\int_{S^2} d \Omega' ~	 \delta(x \cdot {\bf l}) &=  -8 \pi^2  \phi(\mathbf{x})~~, 
		\\
		Q\int_{S^2} d \Omega' ~	\partial_v \delta(x \cdot {\bf l}) &= -8 \pi^2 \partial_v \phi(\mathbf{x}) ~~,
		\\
		Q\int_{S^2} d \Omega' ~	l_v l_i \delta(x \cdot {\bf l}) &=- 4 \pi^2 x^i \partial_v \phi(\mathbf{x})~~, 
		\\
		Q\int_{S^2} d \Omega' ~	l_v^2 \delta(x \cdot {\bf l}) &=-\pi^2 \bigl( \phi(\mathbf{x})+ v \partial_v \phi(\mathbf{x}) \bigr) ~~.				
\end{align}
Substituting these integrals into \eqref{a1} one proves (\ref{3.3cd}).

We now go to derivation of Eq. (\ref{3.46d}). We start with (\ref{3.45}) which one can rewrite as
\begin{equation} \label{a3}
\Phi_s(\Omega,\mathbf{p})\equiv -2\int_{S^2} d \Omega'~ 
\delta(1+(\vec{m}\cdot \vec{e})) \theta (n_v) \Re A({\bf p},{\bf n})~~,
\end{equation}
where ${\bf n}={\bf n}(\Omega')$ is a unit vector whose components can be parametrized as
\begin{equation} \label{a4}
n_v=\cos\theta'~~,~~n_1=\sin\theta'\cos\varphi'~~,~~n_2=\sin\theta'\sin\varphi'~~.
\end{equation}
To proceed with (\ref{a3}) it is convenient to go from the integration on the half of sphere $\vec{n}^2=1$, $n_v>0$ to an integration on the sphere
$\vec{m}^2=1$. By using relation (\ref{3.17}) one concludes that
\begin{equation} \label{a5}
m_x=\cos2\theta'~~,~~m_1=\sin2\theta'\cos\varphi'~~,~~m_2=\sin2\theta'\sin\varphi'~~.
\end{equation}
If $d\Omega''$ is the integration measure on $\vec{m}^2=1$, then
\begin{equation} \label{a6}
d\Omega'={1 \over 4n_v} d\Omega''~~,~~n_v=\sqrt{1 +(\vec{m} \cdot \vec{v}) \over 2}~~,
\end{equation}
where $\vec{v}$ is the coordinate velocity of the shockwave, $v_x=1,v_i=0$.
The integration in \eqref{a3} can be easily performed and yields \eqref{3.46d}.

\newpage
\printbibliography

	\end{document}